אפריל 2017

# מאפייני הנהיגה של קשישים ורמת המודעות שלהם לקשיים שבדרך

ד״ר עדית סולברג

דוח מדעי מסכם

מוגש לקרן מחקרים שליד איגוד חברות הביטוח בישראל

03.9066399
03.9066658
aditso@ariel.ac.il
אריאל, 40700

ד״ר עדית סולברג, המחלקה לכלכלה ומנהל עסקים, אוניברסיטת אריאל

אוניברסיטת אריאל בשומרון ע״ר

## תקציר

שיעור הנהגים הקשישים עלה בשני העשורים האחרונים עם עליית חלקם של הקשישים באוכלוסייה ועליית שיעור הקשישים הנוהגים בתוך אוכלוסיית בני גילם. מחקר זה בוחן את מאפייני הנהיגה של הקשישים ורמת המודעות העצמית שלהם לקשיים שבנהיגה, באמצעות ניתוח נתוני הלמ״ס ובעזרת סקר ייחודי, שכלל סקר העדפות מוצהרות, בו רואיינו 205 נהגים קשישים בגילאי 70 עד 80.

הנהגים הקשישים הם ברובם גברים וזאת למרות גידול חד בשיעור הנשים הנוהגות בקרב הקשישים בעשורים האחרונים, חלק גדול מהם בעלי רישיון גם לרכב כבד. רוב הקשישים נוהגים מדי יום כאשר תדירות הנסיעה נמוכה יותר אצל נשים. שיעור התאונות אותם עושים הנהגים הקשישים נמוך יחסית לצעירים, אך הם נפגעים באופן חמור יותר בתאונות בשל שבריריותם. הקשישים נוטים להחליף את הרכב בתדירות גבוהה יחסית לאוכלוסייה הצעירה יותר. מחצית מהם מצהירים כי הם עדיין "מחליפים גלגל" באופן עצמאי.

המרואיינים הפגינו רמת אופטימיות ובטחון עצמי בנהיגה שאינם מתאימים למאפיינים אובייקטיבים של נהיגת קשישים. סתירה עלתה גם בין יחס הסביבה לנהיגתם ובין תחושתם העצמית וכן בין יחסם של אותם קשישים לקבוצת גילם אותה הם מעריכים ברמת נהיגה נמוכה יותר מזו בה הם מעריכים את עצמם כפרטים. חוסר המתאם בתשובות עלה גם בחלקים אחרים בסקר : שיעור אלה שהודו שעברו תאונות דרכים היה נמוך מזה הצפוי, המרואיינים הקנו חשיבות רבה למערכות הבטיחות ברכב אך לא חשבו שהם עצמם זקוקים לה ורוב הקשישים אינם חושבים שיש איזושהי סיבה שיפסיקו לנהוג למרות קבלת הצעות להפסקת הנהיגה והערות בנושא מהמשפחה ומהסביבה. באמצעות סקר העדפות מוצהרות נבחנה רמת הקושי שהקשישים מייחסים לתנאי הנהיגה. נמצא כי הקשישים חוששים בעיקר מתנאי מזג אויר, מנסיעה בשעות הלילה ומנסיעות ארוכות. החשש מנהיגה בשעות הלילה מתחזק אצל הנשים, המבוגרים יותר ואלה הנוהגים בתדירות נמוכה.

נהיגה היא כלי חשוב בשמירה על עצמאות הקשיש ואיכות חייו ומשום כך נזהרו המרואיינים, באופן מודע או לא מודע, מתיוג כמתקשים בנהיגה. לאור הממצאים, יש לשקול השתת אחריות רבה יותר על המערכת הרפואית. רגישות ואחריות של הסביבה של הקשיש לנושא גם היא בעלת חשיבות רבה ומשום כך יש לפעול לבניית מערך הסברה בנושא לקשיש וסביבתו. כמו כן, יש מקום לשקול מתן רישיונות נהיגה מותנים לקשישים לשעות היום, לתנאי מזג האויר או למרחקי נסיעה קצרים בלבד בהתאם למצבם הרפואי. גמישות זו תאפשר לקשיש מחד לשמור אורך חייו ועצמאותו תקופה ארוכה יותר אך תמזער את הסיכון עבורו ועבור משתמשי הדרך והולכי הרגל במצבים בהם התנאים האמורים מעלים חשש ליכולת הנהיגה שלו.

## 1. הקדמה:

מתחילת המאה העשרים עלתה תוחלת החיים במדינות המפותחות בממוצע בכ-30 שנה: מכ-50 שנות חיים בתחילת המאה שעברה למעל 80 שנים כיום. מאמצע המאה הקודמת עולה תוחלת החיים בעיקר בשל עליית תוחלת החיים המותנה של הקשישים[1] כלומר התארכות החיים בגיל המבוגר שלאחר גיל הפרישה (longevity). בני ה-65 שיכלו לצפות לעוד 11 שנות חיים בממוצע בתחילת המאה העשרים יכולים לצפות כיום בממוצע לעוד 20 שנות חיים.[2]

להתפתחויות אלו השפעות מרחיקות לכת כמעט בכל תחומי החיים וביניהם בתחום התחבורה. באופן ישיר צפוי לעלות גיל הנהגים הקשישים ואחוז הנהגים מקרב אוכלוסיית הקשישים. משום כך יש צורך בהתאמה של כלל הסדרי הנהיגה לתהליך התארכות החיים. להזדקנות האוכלוסייה ישנם גם השלכות עקיפות בתחום התחבורה וביניהם הצורך בהנגשה ובשינוי מסלולי התחבורה הציבורית, הצורך בהתאמת המדרכות ודרכי הגישה לקשישים, שינוי שיטת השילוט ועוד. גם יצרני המכוניות ידרשו להתאים את הרכבים לצרכי האוכלוסייה הקשישה המאופיינת בקשיי נהיגה ובפגיעות פיזית רבה.

העלייה במשקל הנהגים הקשישים המאפיינת מדינות מתקדמות רבות, הובילה להתרחבות הדיון בנושא בספרות המדעית תוך בחינת השפעת עליית משקלם של הנהגים הקשישים על הסדרי הנהיגה, דפוסי הנהיגה שלהם, היקף ועוצמת תאונות הדרכים ועוד. יש המתייחסים לנהגים הקשישים כנהגים פחות טובים בשל ירידה ביכולות הפיסיות והמנטאליות המלוות פעמים רבות את הזקנה. מנגד, יש הטוענים כי דווקא הנהגים הקשישים זהירים יותר ונוהגים באופן בטוח יותר. בין התוצאות החזקות העולות מהספרות העולמית נמצאת העובדה כי שיעור תאונות הדרכים של הקשישים נמוך יחסית לגודל אוכלוסיית הנהגים הקשישים ולהיקף נהיגתם, ממצא הנקשר לעובדה שהקשישים נוהגים בעיקר בשעות היום, בנסיעה איטית עירונית ולמרחקים קצרים. עם זאת, חומרת הפגיעה שלהם בתאונות גדולה יותר.

הספרות מתייחסת גם לקשיים הפיזיים, הטכניים והמנטליים הכרוכים לנהיגת קשישים ובוחנת את מנגנון הניטור והוויסות העצמי של הקשישים. בעוד שחלק גדול מהמאמרים טוען כי מנגנון זה טוב יותר משל הצעירים, וקשישים בעלי ירידה ביכולות הפיסיות, הטכניות או המנטליות ימנעו באופן עצמאי מלנהוג, מסתפקים מאמרים אחרים, האם מנגנון זה מספק, או שמא יש צורך בהתערבות רגולטורית ובהגבלה חוקית נוספת של נהיגת הקשישים.

בישראל עלה בתוך עשור חלקם של הנהגים מעל גיל 65 מסך הנהגים מ-6.9% ל-10.3% (שנים 2005 ו-2015 בהתאמה). עלייה זו מקורה בשתי התפתחויות משמעותיות: עליית חלקם של הקשישים באוכלוסייה ועליית חלקם של הנהגים בקרב אוכלוסיית הקשישים. במחקר זה ייבחנו דפוסי הנהיגה של הקשישים בישראל ורמת המודעות העצמית שלהם ביחס לאיכות נהיגתם. במהלך המחקר בוצע סקר בו נדגמו ונשאלו 205 נהגים קשישים בגילאים 70 עד 80 מגוון רחב של שאלות בנושאים אלו. הסקר אפשר להציג מאפיינים נוספים של נהיגת הקשישים שלא זוהו עד היום וכן לבחון את סוגיית המודעות והניטור העצמי מזוויות מבט רבות ושונות.

כמחצית מהקשישים נוהגים, רובם גברים. מאפיינים מיוחדים נוספים כמו החזקת רישיון במשך עשרות שנים, נהיגה בתדירות גבוהה יחסית, שיעור גבוה של בעלי רישיונות לרכב כבד משאית או מונית, שיעור נמוך יחסית של תאונות וכן

---

[1] על פי ההגדרה הסטטיסטית המקובלת קשישים מוגדרים כאלה שגילם מעל 65 שנה.
[2] תוחלת חיים מותנה היא תוחלת החיים שיש לפרט בהינתן שהגיע לגיל מסוים. כך לדוגמא, לפי נתוני הלמ״ס תוחלת החיים המותנה של גבר בגיל 65 היא 19 שנים ושל אשה בגיל 65 21.3 שנה. בגיל 80 תוחלת החיים המותנה היא 8.7 שנים ו-9.6 שנים לגברים ונשים בהתאמה (נתוני למ״ס 2014).

תכיפות גבוהה יחסית של החלפת מכונית ושמירה על עצמאות בתיקונים לרכב מגדירים את הנהגים הקשישים כקבוצה. כמו כן נמצא כי יותר ממחציתם מחויבים בהרכבת משקפיים וכי שיעור הקשישים הנדרשים באישור מיוחד בשל מחלה הוא גבוה משמעותית מזה של הצעירים.

רמת המודעות והניטור העצמי של הקשישים ביחס לנהיגתם נבחנו ממספר זוויות. השאלות התייחסו להערות בני המשפחה בדבר נהיגת הקשיש, תחושתו העצמית, הערכתו לגבי נהיגת קשישים אחרים ומידת הסכמתו להיגדים שונים שנאמרו בנושא. התשובות שהתקבלו מציגות תמונה מורכבת מאוד המצביעה על רצונו של הנהג הקשיש להימנע מלקשר עצמו לאיזשהו קושי או מגבלה הקשורים לנהיגת קשישים. המידור העצמי של הנהג מהחולשות הכרוכות בנהיגת קשישים, לא מונע ממנו להכיר בחולשות אלו אצל נהגים קשישים אחרים או להבין כי בני משפחתו והסביבה מעריכים את נהיגתו בצורה שונה, בעייתית יותר.

בשלב האחרון של המחקר בוצע סקר העדפות בו נתבקשו הנהגים הקשישים שנדגמו לבחור בין שתי אופציות נהיגה שונות המורכבות ממספר מאפיינים. לדוגמא בין נהיגה בלילה, בכביש עירוני, במזג אויר חורפי, ללא מלווה ולמרחק גדול ובין נסיעה ביום, בכביש עירוני, במזג אויר חורפי, עם מלווה ולמרחק קטן. כל מרואיין נדרש לבחור את האופציות העדיפות עליו מבין חמישה צמדים שונים. באמצעות אמידת התועלת מכל אחד ממאפייני הנהיגה, אופיינו הנסיעה במזג אויר גרוע וכן הנסיעה בלילה כמגבלות אותה רואים הקשישים כבעייתיות ביותר. החשיבות היחסית של מגבלות הדרך משתנה בין קבוצות קשישים שונות בסיווג לפי גיל, מגדר, השכלה ועוד.

נהיגה היא כלי מרכזי בהגדרת העצמאות של הקשיש ובתחושת החופש שלו. אובדן היכולת לנהוג והניידות העצמאית, יכולים לגרום לירידה משמעותית באיכות חיו של הקשיש ובתחושת הערך העצמי שלו. עם התארכות משך החיים של הקשיש ובהינתן הגישה הרווחת היום כי יש לדאוג ככל הניתן לחיים עצמאיים ומלאים של הקשיש במסגרת הקהילתית, אובדן כלי זה על כל הכרוך בו, תפקודית או נפשית, נעשה משמעותי יותר.

התמונה המורכבת העולה במאמר קשורה כפי הנראה לחשיבות הרבה שמקנים הקשישים לנושא הנהיגה כסמן לעצמאות וכמשפיע מהותית על איכות חייהם. קשר זה מוביל לאופטימיות יתר שלהם ביחס לנהיגתם ולפער בין יכולות הנהיגה בהם הם מכירים ובין יכולות הנהיגה בפועל. פער זה מחייב התייחסות שכן מחד חשוב לשמור על עצמאות הנהיגה של הקשיש ככל שניתן, אך במקביל, יש למנוע נהיגת קשישים כשהיא אינה בטוחה דיה לקשיש עצמו, למשתמשי הדרך ולהולכי הרגל.

בחלק הבא תסקר הספרות האקדמית הנוגעת לנושאים בהם דן המאמר. בחלק השלישי יבחנו הרגלי הנהיגה של הקשישים כפי שהם עולים מסקרי הלשכה המרכזית לסטטיסטיקה. בחלק הרביעי יוצגו הרגלי הנהיגה שהתקבלו מהסקר הייחודי שנערך במחקר זה. הרגלי הנהיגה העולים בסקר יבחנו מול התוצאות העולות מסקרי הלמ"ס וכן יוספו מאפיינים נוספים על אלו שנבחנו עד היום. החלק החמישי יבחן את רמת המודעות והניטור העצמי של קשישים ביחס לנהיגתם. החלק השישי יבדוק באמצעות מודל העדפות את הרגלי הנהיגה ואת הקשיים מהם חוששים הקשישים. אפיונים אלו יעזרו להגדיר דרכים בהם ניתן להאריך את תקופת הנהיגה של הקשיש תוך הקטנת הקושי בנהיגה והסכנות הכרוכים בהם. דרכים אלו יבחנו בחלק השביעי והמסכם של העבודה.

## 2. סקירת ספרות

מספר מחקרים בדקו את הקשר בין יכולות נהיגה וגיל מבוגר: .Adrian et al (2011) בדקו את השפעת פונקציות תפקודיות ואישיות בקרב נהגים קשישים ומצאו כי מין וגיל הם המנבאים הטובים ביותר ליכולת נהיגה, כאשר איכות הנהיגה יורדת עם עליית הגיל ואצל נשים. .Selander et al (2011) ערכו מבחנים לנהגים קשישים ומצאו כי אין קשר בין הערכה העצמית של נהג ובין איכות נהיגתו. כמו כן הם מצאו כי אחוז גבוה מהקשישים נכשלים במבחני הנהיגה. Tefft (2008) בחן את מאפייני הנהגים הפוגעים בתאונות דרכים וייחס סיכוי גבוה יותר בגרימת תאונות דרכים לנהגים קשישים. Bao and Boyle (2009) הראו כי נהגים קשישים זהירים פחות בצמתים ותנאי שטח מורכבים ומשום כך נפגעים יותר במקומות אלו. גם Pollatsek et al. (2012) וכן Romoser and Fisher (2009) מצאו כי נהגים קשישים נוטים להיפגע יותר בצמתים. לטענתם, הדבר קשור להרגלי חוסר זהירות וסריקת הכביש ופחות לבעיות קוגניטיביות או פיזיות.

Bunce et al. (2012) בחנו את היכולות של קשישים להאיץ, לשמור מרחק ולעבור נתיב באמצעות סימולטור. הם הראו שנהגים קשישים מפגינים חוסר עקביות גבוה בביצועי הנהיגה שלהם כנראה בשל הקושי בשמירה על קשב בקרב הקשישים. Musselwhite and Haddad (2010) הראו כי למרות שנהגים קשישים מעריכים את עצמם כנהגים טובים יותר מנהגים צעירים, הם מדווחים על קושי בקריאת תמרורים, בשמירה על מהירות קבועה, על רגישות לבוהק מן שמש ועוד. Middleton et al. (2005) בדקו את משך הזמן הדרוש בקבלת החלטות בנהיגה ואילו Herriotts (2005) בחנו את היכולת להפנות מבט לאחור ולהכנס ולצאת מהמכונית. בשני המאמרים נמצא כי נהגים קשישים מתקשים יותר בפעולות אלו.

בניגוד לספרות המקשרת בין העלייה במספר הקשישים לנהיגה איכותית פחות מוצג בספרות גם הקשר ההפוך: .Smith et al (1993) הראו כי השונות בקרב הקשישים היא הגדולה ביותר בין כל קבוצת גיל, ולכן, גיל כרונולוגי הוא אינדיקטור חלש להיבטים ההתנהגותיים של נהיגה. .Borowsky et al (2010) הראו כי גיל מתקדם כמעט אינו משפיע על יכולתם של נהגים מבוגרים לתפוס סיכון וכי הם מודעים לבעיות הקשורות לגיל. .Langford et al (2006) טענו כי כאשר בודקים את שיעור התאונות יחסית לנסועה, נהגים קשישים נפגעים בשיעור נמוך יותר של תאונות דרכים. Alvarez and Fierro (2008) טענו כי נהגים קשישים טובים לא פחות מנהגים צעירים. הם הראו שבעיות רפואיות או אחרות גורמות להם לשיעור נסועה נמוך או להפסקת נהיגה ומשום כך אלו שנותרים בכביש הם הנהגים הטובים והזהירים. Davey and Nimmo (2003) הראו שיעדי הנהיגה של קשישים שונים מאלה של נהגים צעירים, היעדים עירוניים וקרובים יותר, ומשום כך נהיגתם מסוכנת פחות. בדומה .Taubman et al (2004) ו-.Horswill et al (2011) אפיינו את הנהיגה של המבוגרים כנהיגה בטוחה ושקולה יותר.

מספר רב של מאמרים בדק את הקשר בין גיל ובין מעורבות ורמת הפגיעות מתאונות הדרכים. .Dellinger et al (2004) בחנו את רמת ההיפגעות בתאונות דרכים והגיעו למסקנה כי תוצאות תאונות הדרכים בהם מעורבים צעירים קטלניות יותר מאשר תאונות דרכים בהם מעורבים קשישים. בנוגע לרמת הפגיעות מתאונות של הנהגים הקשישים הראו .Li et al (2003) כי שיעורי תמותה בתאונות דרכים גבוהים יותר אצל מבוגרים מגיל 60-64 ועולים מונוטונית עם הגיל. .McCoy et al (1989) הראו שלא רק ששיעור התמותה עולה בגיל נהגים מבוגרים אלא גם הפגיעות לכל סוג תאונה קשות יותר בקרב המבוגרים. Abd-Raya and ElMeguid(2009) בדקו את הסיבות והמאפיינים של תאונות דרכים במטרה לפתח אסטרטגיות למניעתן.

הם מצאו כי תאונות בהן מעורבים אנשים מבוגרים נפוצות בעיקר בקרב גברים, בצמתים ובתאונות אלה פציעות החזה ובפרט השברים בצלעות היו נפוצות. לטענתן אפיון שבריריות של נהגים מבוגרים צריך להילקח בחשבון ומשום כך יש ליישם אסטרטגיות לבטיחות כגון בדיקות רפואיות תקופתיות המיועדת לאוכלוסייה הקשישה.

הקשר בין קושי בריכוז וראיה ובין קשיים בנהיגה נבדק גם בקרב חוקרים מתחום הרפואה. Anstey et al. (2005) סקרו מספר רב של מאמרים רפואיים הקושרים בין קשיים קוגניטיביים ופיסיים ורמת היפגעותם בתאונות דרכים. הם מציעים מבחנים לאיתור קשיים אך לטענתם רק ניטור עצמי על ידי הקשיש תהיה לו השפעה משמעותית בהפחתת תאונות הדרכים. הקשר בין נהגים מבוגרים, ניטור עצמי (או הסדרה עצמית) שיקול הדעת ורמת הנהיגה שלהם נבחן על ידי מאמרים רבים נוספים כאשר החוקרים נחלקים לגבי עוצמת הניטור העצמי והצורך בהסדרה. Hakamies-Blomqist and Wahlstrom (1998) הראו כי מרבית הנהגים בני ה-70 שהפסיקו לנהוג עשו זאת בהחלטה עצמית ולא בשל הנחייה שקיבלו. הם הראו כי נהגים שהפסיקו לנהוג היו במצב בריאותי פחות טוב מהנהגים וכי ככל שהקשישים הפסיקו לנהוג בגיל צעיר יותר מצב הבריאותי היה פחות טוב יחסית לבני גילם. Mend and Siren (2012) הראו קשר לינארי בין אי נוחות בנהיגה והימנעות מנהיגה. מאחר ונמצא כי אי הנוחות קשורה גם לירידה בתפקודים הרי ירידה בתפקוד במיוחד קוגניטיבי מובילה להקטנת הנהיגה של הפרט. Gwyther and Holland (2012) מצאו כי רמת הניטור העצמי עולה עם הגיל כמו גם עם רמת החרדה וחוסר הביטחון. Cantin et al. (2009) בדקו את מנגנוני הפיצוי של הנהגים המבוגרים ומצאו כי הם כמנגנון מפצה על ירידת היכולת הם נוהגים לאט יותר ומשום כך בזהירות רבה יותר. לעומת טענו Bell et al. (2015) כי חוקים מקשים כמו זמן חידוש רישיון מקוצר ובדיקות רפואיות הם אלו המשפיעים על שיעורי תמותה נמוכה יותר בתאונות בקרב מבוגרים. ואילו Langford et al (2008) כי החלת חוקים מקשים יותר על נהגים קשישים לא פועלת להורדת שיעור התאונות בקרבם. Ross et al. (2009) הראו כי הניטור העצמי של הנהגים אינו מספיק וכי הם בסיכון כפול מצעירים לכשלים בנהיגה.

Brayne et al. (2000) בחנו את מאפייני אוכלוסיית הקשישים הנוהגת מעל גיל 85 (very old) והראו כי בגיל המבוגר – מעל 85 – עדיין כ-10% נוהגים, רובם באופן קבוע. רוב הנהגים הם גברים נשואים. מבין אלו שוויתרו על נהיגה, כמחצית ויתרו בגיל 80 ומעלה. כשליש ויתרו על הנהיגה בשל בעיות בריאות, כשליש ויתרו בשל עצה שקיבלו ו-20% נוספים ויתרו בשל חוסר בטחון עצמי. ניטור עצמי הוביל לכך שהנוהגים בגיל המבוגר הם בעלי כושר תפקוד גופני ומנטלי גבוה יחסית לקבוצות גיל אחרות. הם הניחו שבדיקות רפואיות לזיהוי דמנציה או מצב בריאותי באוכלוסייה המבוגרת לא יפחיתו את העלייה בתאונות בקרב הקשישים שכן נהגים אלו בוחרים להפסיק לנהוג באופן עצמאי. Robertson (2008) בחן את דעת הציבור ביחס לנהגים מבוגרים ומצא כי בקרב האוכלוסייה הכללית ישנן רמות גבוהות של דאגה בנוגע לנהגים מבוגרים. רמת הדאגה ביחס לנהיגת קשישים פוחתת עם עליית הגיל של הנשאל.

סוג אחר של מחקרים דן ביכולת של הנהגים המבוגרים להשתמש באמצעי ניווט ותקשורת במכוניות. Emmerson et al. (2013) מצאו כי מכשירי הניווט מקילים על הנהג המבוגר ונותנים לו תחושת ביטחון. לעומת זאת Fofanova and Vollrath (2011) מצאו כי הנהג המבוגר מתקשה לבצע מספר משימות בו זמנית ומשום כך מתקשה בתפעול מערכות אלו ובזמן הנסיעה דבר העלול לפגוע בבטיחות נסיעתו.

## 3. מאפייני נהיגת הקשישים בישראל (נתוני למ"ס)

### 3.1 אפיון הקשישים הנוהגים

מספר מחזיקי רישיון נהיגה ישראלי בתוקף בישראל עמד בשנת 2015 על 4.02 מיליון מתוכם כ-440 אלף היו נהגים מעל גיל 65 (10.9% מכל הנהגים) כאשר 120 אלף מהם היו מעל גיל 75 (3% מכל הנהגים)[3]. בקרב בני ה-25 עד 54 כ-80% מחזיקים ברישיון נהיגה ואילו בקבוצת הגיל 65 ומעלה רק כ-48% מחזיקים ברישיון נהיגה. שיעור זה גבוה יחסית ועומד על 63.5% אצל אלה שרק יצאו לפנסיה, קבוצת הגיל 65-74, ויורד לפחות ממחצית ולשיעור של 28.8% בקרב אלה שגילם מעל 75. מספר הנהגים הקשישים ושיעורם מכלל הנהגים עלו בשני העשורים האחרונים חדות (ראו דיאגרמה 3.1). עלייה זו מקורה בשתי התפתחויות משמעותיות: עליית חלקם של הקשישים באוכלוסייה ועליית חלקם של הנהגים בקרב אוכלוסיית הקשישים.

**דיאגרמה 3.1: שיעור הנהגים הקשישים מסך הנהגים בישראל, 2003-2015 בחלוקה לפי גיל, אחוזים**

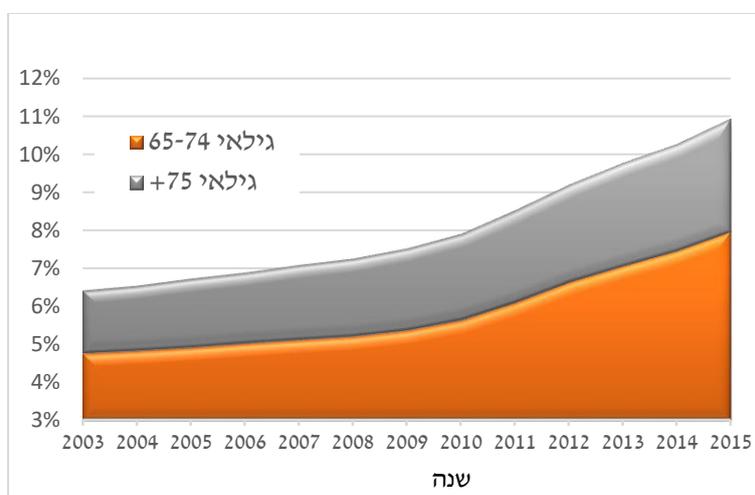

בעשור האחרון עלה חלקם של הקשישים באוכלוסייה ל-11% מהאוכלוסייה כתוצאה של הגעת שנתונים גדולים לגיל פרישה ועליית תוחלת החיים המותנה בקרב הקשישים. תחזיות שונות לגבי משקל הקשישים באוכלוסייה ותוחלת החיים המותנה מצביעות על ההאצה נוספת הצפויה במשתנים אלו (Sohlberg and Yutav, 2017). עם זאת, שיעור הנהגים הקשישים מסך הנהגים עלה בשיעור גבוה יותר משיעור עליית חלקם באוכלוסייה. זאת, בשל עליית שיעור הנהגים בקרב הקשישים עצמם. שיעור מחזיקי רישיון הנהיגה בקרב הקשישים עלה במהלך השנים יותר מבכל קבוצה אחרת: משנת 1995 ל-2015 עלה שיעור הנהגים מ-21.5% ל-48.2% בהתאמה, כלומר הכפיל את עצמו ויותר בתוך עשרים שנה (דיאגרמה 3.2). בתוך אוכלוסיית הקשישים עלה שיעור מחזיקי הנהיגה בשתי קבוצות הגיל: בקבוצת הגיל 65 עד 74 הוא עלה מ-43.1% בשנת 2005 ל-63.5% בשנת 2015 ובקרב הקשישים שגילם מעל 75 הוא עלה מ-19% בשנת 2005 ל-28.8% בשנת 2015. כל אלה הביאו לעליית במשקל הקשישים המורשים לנהוג ולעלייה בגילם הממוצע של הקשישים[4].

---

[3] כל הנתונים בחלק זה הם נתוני הלמ"ס לשנים הרלוונטיות.
[4] אם נניח כי שיעור הנהגים בני +75 היה 20% משיעור הנהגים הקשישים בשנת 1995 עלה הגיל הממוצע של הנהגים הקשישים מ-39 שנה ב-1995 ל-43 שנה ב-2015.

**דיאגרמה 3.2 : שיעור מחזיקי רישיון נהיגה מתוך סך האוכלוסייה בגיל, שנים נבחרות, אחוזים**

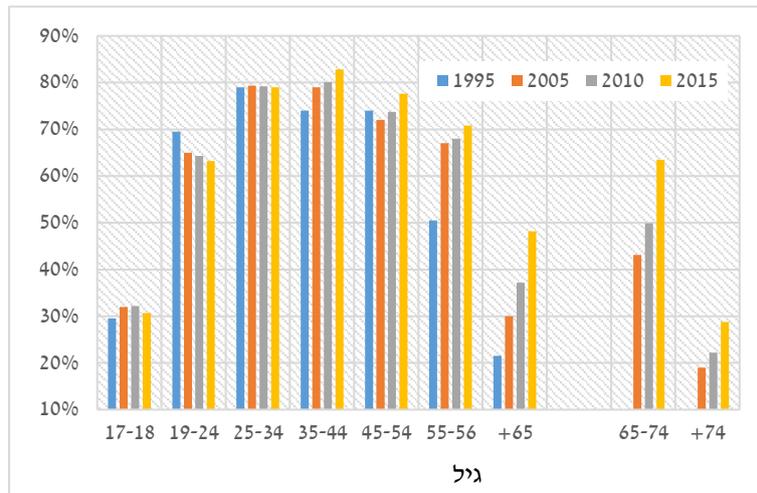

**דיאגרמה 3.3 : שיעור מחזיקי רישיון נהיגה בחלוקה לפי מגדר וקבוצת גיל, 2015, אחוזים**

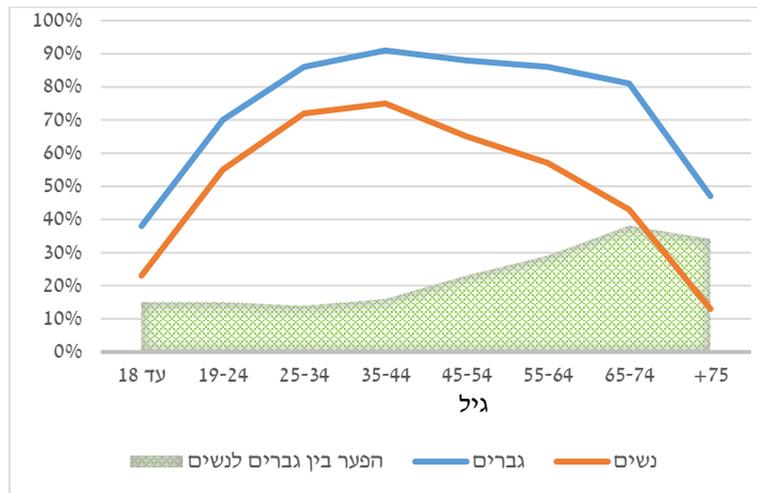

מכלל המורשים לנהוג בני 65 ומעלה היוו הגברים 64%, שיעור גבוה יחסית לחלקם באוכלוסייה בקבוצת גיל זו המוטה לכיוון הנשים (מקרב בני ה-65 ומעלה 44% הם גברים ו-56% נשים). הטיה זו לכיוון הנהגים הגברים מאפיינת את קבוצות הגיל אך היא הולך ומתרחבת עם עליית השנתונים, מ-14 נקודות אחוז בקרב הצעירים בני העשרים לפער של 38 נקודות אחוז בקרב בני 65 עד 74. נראה כי על אף היחלשות המוסכמות והסטיגמות, תפקיד ה״נהג״ נשאר באופן מסורתי אצל הגבר. בקרב קבוצת הקשישים הטיה זו לכיוון הנהגים הגברים חזקה במיוחד ונובעת מהנורמות שהיו מקובלות בעבר ומשיעור הנשים הנוהגות הנמוך שאפיין את השנתונים המבוגרים. יחד עם זאת, על אף הגידול המשמעותי בשיעור הגברים הנוהגים בעשור האחרון – גידול של 85% בין השנים 2005 ל-2015 - שיעורם של הנשים הקשישות בעלות רישיון נהיגה עלה בשיעור מדהים אך יותר של 175%.

שיעור מחזיקי המכוניות בקרב הקשישים נמוך מזה של האוכלוסייה הצעירה יותר – 55.1% לעומת 72.1% בהתאמה, דבר המשקף את שיעור הנהגים הנמוך יותר בקרבם יחסית לאוכלוסיות הצעירות יותר (לוח 3.1). כצפוי שיעור זה גבוה אצל בני

ה-65-74 ומתקרב לשיעור המאפיין את האוכלוסייה הצעירה יותר ואילו בקרב הקשישים בגילאים המבוגרים יותר (75+) שיעור זה יורד ל-38% בלבד. עם זאת, גם שיעור מחזיקי הרכב בקרב הקשישים עלה משמעותית בעשור האחרון עם עליית שיעור הנהגים בקרב הקשישים. רק ל-13.8% מהקשישים יש יותר ממכונית אחת לעומת שיעור כפול אצל האוכלוסייה הצעירה יותר.

**לוח 3.1 מאפייני הנהגים הקשישים לפי מגדר, 2005 ו-2015, נתוני הלמ״ס**

| | 2015 | | | 2005 | | | |
|---|---|---|---|---|---|---|---|
| נשים | גברים | סה״כ | נשים | גברים | סה״כ | | |
| 158.8 | 280.5 | 439.3 | 57.7 | 151.8 | 209.5 | מספר הנהגים הקשישים (אלפים) |
| 8.4% | 11.6% | 10.3% | 6.7% | 8.9% | 6.9% | שיעור הנהגים הקשישים מסך הנהגים |
| 28.2% | 65.3% | 44.6% | 14.5% | 51.3% | 30.2% | שיעור הנהגים בקרב קבוצת הקשישים |
| | | 36.2% | | | 27.5% | שיעור הנשים הנוהגות |
| 42% | 78.6% | 63.5% | 22.2% | 64.1% | 43.1% | שיעור הנהגים בקרב גילאי 65-74 |
| 13% | 47.1% | 28.8% | 6.3% | 34.7% | 19.0% | שיעור הנהגים בקרב גילאי 74+ |
| | | 29.1% | | | 21.2% | שיעור הנשים הנוהגות בקרב בני 75+ |
| 49.2% | 62.8% | 55.1% | 27.7% | 44.1% | 34.8% | שיעור מחזיקי הרכב בקרב הקשישים |
| 175.2% | 84.7% | 109.6% | | | | שיעור הגידול במספר הנהגים בשנים 2005 עד 2015 |
| 0.1% | 21.7% | 13.9% | חוסר בנתונים | | | שיעור המחזיקים רישיון על רכב כבד מקרב גילאי 65+* |
| 0.1% | 12.5% | 7.8% | חוסר בנתונים | | | שיעור המחזיקים רישיון על מונית או אוטובוס מקרב גילאי 65+ |
| 0.3% | 0.6% | 0.5% | חוסר בנתונים | | 0.7% | שיעור הקשישים המעורבים בתאונות מסך הקשישים הנוהגים (בשנה) |
| 4.6% | 9.3% | 7.7% | חוסר בנתונים | | | שיעור המורשעים מעברות תנועה מסך הנהגים בגיל |

* רישיון לרכב כבד וגרר.

רוב מוחלט של הקשישים (מעל 95%) הם בעלי רישיון נהיגה בתוקף יותר מ-15 שנה. כאשר בוחנים את סוג רישיונות הנהיגה של הקשישים עולה תמונה מעניינת: הקשישים מחזיקים בשיעור גבוה יחסית לאוכלוסיית הצעירים גם ברישיונות לנהיגת משאית ובפרט משאית כבדה, מוניות, אוטובוסים או גוררים (ראו דיאגרמה 3.4). גם שיעור הרישיונות לרכב דו גלגלי (אופנוע) גבוה בקרב הקשישים. שיעור הנהגים בעלי רישיונות מיוחדים היה גבוה יותר באמצע המאה הקודמת בעיקר על רכבים כבדים וירד לאחר מכן בשיעור ניכר (ראו דיאגרמה 3.5). כך לדוגמא שיעור בעלי רישיון על רכב כבד (משאית) מכלל הנהגים ירד מ-31.2% אחוזים בשנת 1970 ל-11.8% אחוזים בלבד בשנת 2014. תופעה זו מוסברת בקלות קבלת הרישיונות המיוחדים בעבר, בצורך בנהגים הן לצרכים צבאיים והן בשל מפעלי הבניה הגדולים שהיו אז ושאופיינו חלקם במרחק רב (סיני) ולהתמחותה של ישראל בעשורים הבאים בענפים טכנולוגיים הדורשים כוח עבודה שהוא פחות המיומן בנהיגה והובלה מקצועיים. ייתכן גם כי מדובר בהטיה של המדגם שכן נהגים בעלי רישיון מיוחד הם יותר מיומנים ומחוברים לנהיגה ומשום כך ייפרדו כנראה מ״ההגה״ בגיל יותר מבוגר, אם בשל מיומנותם ואם בשל ״החיבורי״ שלהם לנהיגה. שיעור הנהגים

"הכבדים" יורד לאחר גיל 75, אך עדין גבוה יחסית אוכלוסייה גבוהה כנראה בשל הדרישות הרפואיות והטרחה הכרוכה בהם הגורמים לאוכלוסייה המאוד מבוגרת להחזיר את הרישיונות המיוחדים.

**דיאגרמה 3.4 : שיעור בעלי רישיונות מיוחדים בנוסף על רישיון הנהיגה על רכב פרטי לפי גיל, 2015, אחוזים**

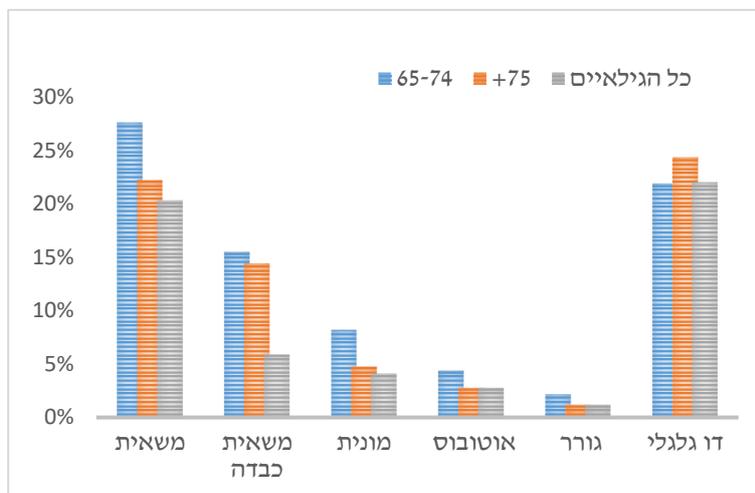

**דיאגרמה 3.5 : שיעור בעלי רישיונות מיוחדים מכלל הנהגים, 1970-2010, אחוזים**

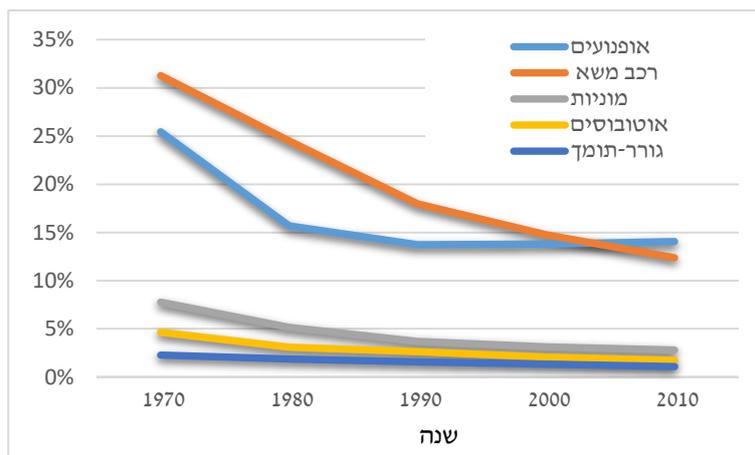

בבחינת המגבלות החלות על הנהג נמצא כי כצפוי, שיעור המגבלות עולה עם הגיל: שיעור הנהגים המחויבים להרכיב משקפיים או עדשות מגע נשאר קבוע על כ-25% מהנהגים עד גיל 54, עולה מעט בקבוצת מגיל 55 ואז עולה חדות בקרב הקשישים שגילם מעל 65, עם עלייה נוספת בגילאי 75+. כ-60% מקרב הנהגים בגילאי 65-74 וכ-70% מקרב הנהגים בגילאים שמעל 75 מחויבים בהרכבת משקפים או עדשות בנהיגה (דיאגרמה 3.6). גם שיעור הנהגים המחויבים בבדיקות רפואיות בשל מגבלה רפואי עולה ומגיע ל-10% בקרב גילאי 65-74 ול-25% בגילאים 75+ לעומת שיעורים נמוכים בהרבה בקבוצות הגיל הצעירות יותר.

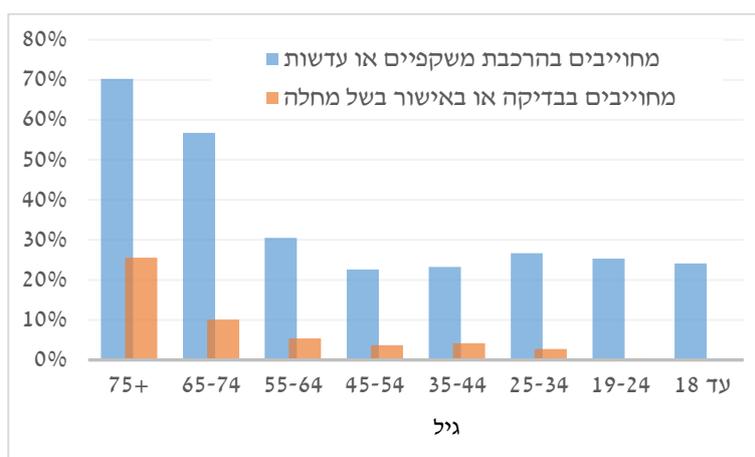

**דיאגרמה 3.6: שיעור הנהגים המחויבים בהרכבת משקפיים או בבדיקות רפואיות בשל מצב בריאותי מקרב הנהגים באותה קבוצת גיל, 2015, אחוזים**

### 3.2 תאונות דרכים עם נפגעים בקרב קשישים – נתוני הלמ"ס

למרות הגידול באוכלוסייה, במצבת הרכב ובנסועה בעשור האחרון, קיימת מגמת ירידה במספר המוחלט של התאונות עם נפגעים בכלל האוכלוסייה: מ-32.3 אלף תאונות בשנת 2003 ל-21.8 אלף תאונות בשנת 2015. שיפור זה הוביל לירידה מתמשכת באחוז הנהגים המעורבים בתאונות מקרב הנהגים הן באוכלוסיית הקשישים והן בכל האוכלוסייה (דיאגרמה 3.7). עם זאת, העלייה בשיעור הקשישים באוכלוסייה ובפרט בשיעור הנהגים הקשישים הובילה להכפלת שיעור תאונות הדרכים בהם מעורב קשיש מכלל התאונות תוך עשור (דיאגרמה 3.8). בקרב בני ה-85+ שיעור התאונות מכלל התאונות אפילו שולש בשל הגידול החד בחלקה של קבוצה זו בכלל הנהגים. ראוי לציין כי נתונים אלו כוללים גם את הנפגעים המעורבים בתאונה כהולכי רגל שחלקם גבוה יחסית בקבוצת הקשישים (כ-35% מהקשישים הנפגעים בתאונות דרכים היו הולכי רגל לעומת 14% בכלל האוכלוסייה).

אחוז הקשישים המורשעים בעברות תאונה נמוך מזה של הצעירים. באופן דומה, שיעור תאונות הדרכים בקרב האוכלוסייה הקשישה כולה נמוך יחסית לחלקם של הקשישים באוכלוסייה והוא נמוך מ-10%. השיעור הנמוך של הנהגים הקשישים בקרב בני גילם ושיעור התאונות שמבצעים נהגים קשישים, שהוא נמוך יותר מזה של הצעירים מהם, מסבירים את חלקם הנמוך של הקשישים בסך התאונות (דיאגרמה 3.9). משנת 2011 רואים שתי מגמות הפוכות – בעוד ששיעור הנהגים הקשישים המעורבים בתאונות קטלניות יורד (מלבד שנת 2015 שהייתה חריגה לרעה בנושא זה), שיעור הקשישים המעורבים בתאונות קלות ממשיך לעלות. התפלגות התאונות בקרב הקשישים דומה לזו של הצעירים יותר כאשר הפגיעה הנפוצה ביותר היא פגישת חזית-צד ואחריה פגיעה בהולך רגל.

**דיאגרמה 3.7: שיעור הנהגים המעורבים בתאונות דרכים מתוך סך האוכלוסייה בגיל, 2006-2015, אחוזים**

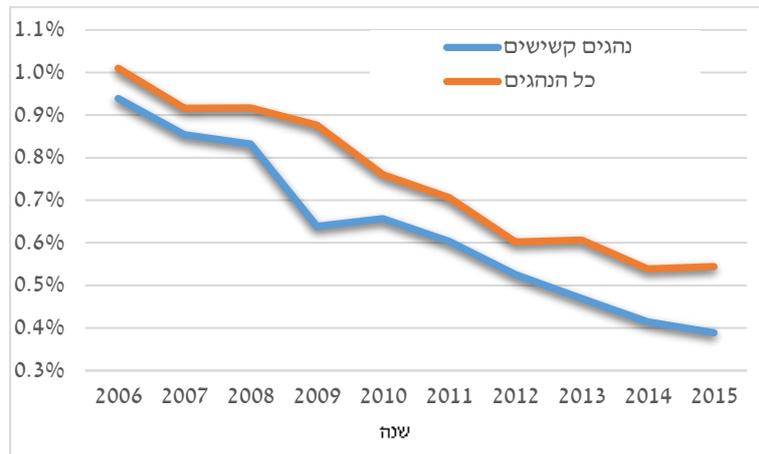

**דיאגרמה 3.8: שיעור הנפגעים וההרוגים הקשישים שהיו מעורבים בתאונות דרכים מתוך סך המעורבים בתאונות דרכים, 2006-2015, אחוזים**

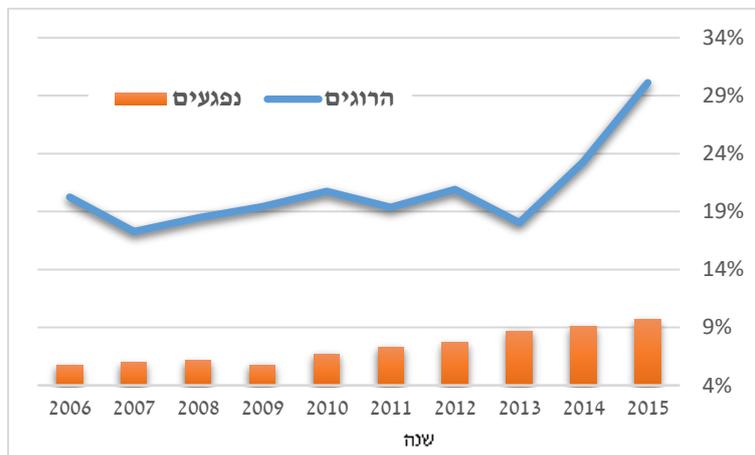

**דיאגרמה 3.9: שיעור תאונות הדרכים בהם מעורב נהג קשיש מתוך סך תאונות הדרכים בסיווג לפי עוצמת התאונה, 2006-2015, אחוזים**

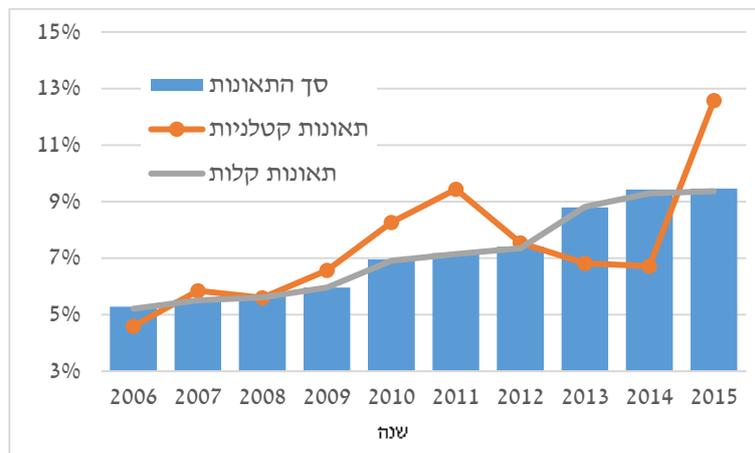

מספר הקשישים שנהרגו בתאונות דרכים ירד בעשור האחרון במקביל לירידה במספר ההרוגים הכלל בתאונות דרכים (מלבד בשנת 2015 בה חלה עלייה חדה בתאונות ובנפגעים). עם זאת, שיעור הקשישים הנהרגים ונפגעים בתאונות דרכים עולה לאורך זמן עם הגידול בשיעור הקשישים הנוהגים אחוז הקשישים ההרוגים בתאונות דרכים גבוה ברוב השנים בעשור האחרון פי 2 (ובשנת 2015 כמעט פי 3) משיעורם היחסי באוכלוסייה. ראוי לציין כי בעוד רוב הנפגעים בקרב האוכלוסייה הצעירה (עד גיל 65) הם הנהגים, בקרב הקשישים התמונה שונה ומשקלם של הנפגעים שהם הולכי רגל דומה לזה של הנפגעים בקרב הנהגים. הקבוצה השלישית בגודלה בין הנפגעים הקשישים היא קבוצת הנוסעים.

## 4 הרגלי הנהיגה בקרב קשישים – סקר מיוחד

### 4.1 תיאור הסקר

במסגרת המחקר נערך סקר מיוחד שהתמקד בהרגלי הנהיגה של קשישים בקרב 205 קשישים בעלי רישיון נהיגה בגילאי 70 עד 80. הסקר נערך בחודשים אוקטובר ונובמבר 2016 באזור חיפה והקריות בצפון ובתל אביב, רמת גן ופתח תקווה במרכז. הראיונות התקיימו במקומות מפגש של הקשישים: מתנ"סים בשעות אחר הצהריים, בתי קפה וקניונים בשעות הבוקר וכן תוך כדי נסיעה ברכבת במהלך כל היום. כל ראיון נמשך כחצי שעה ונעשה על ידי סוקרת מבוגרת המיומנת בביצוע ראיונות מסוג זה. סקרים מסוג "העדפות מוצהרות" הם ארוכים יותר ודורשים הרבה סבלנות, משום כך מקומות המפגש שנבחרו הם כאלה בהם קל יותר לראיין את האנשים בניחותא. .

בשאלון היו כ-40 שאלות שחולקו ל-6 חלקים (השאלון המלא מצורף בנספח 1) :

1. חלק א': שאלות כלליות שמטרתן לבחון האם המרואיין עונה על הגדרות המחקר : מגיל 70 עד 80 עם רישיון נהיגה בתוקף.
2. חלק ב': שאלות סוציו דמוגרפיות : מין, גיל השכלה, עבודה וסטטוס התנדבות.
3. חלק ג': מאפייני כלי הרכב: שנתון, מערכות המותקנות ברכב ותדירות החלפת כלי הרכב.
4. חלק ד': תאונות דרכים: האם נפגע מתאונות דרכים בעבר, מאפייני התאונה (מקום, סוג התאונה, נפגעים וסוג הדרך), טיפול ברכב.
5. חלק ה': הערכת המרואיין את נהיגתו ונהיגת קשישים אחרים : דעה על הגבלת הנהיגה למבוגרים, דעה של הילדים על התנהגות הנהיגה של המרואיין, בחינת הסכמת המרואיין להיגדים שונים.
6. חלק ו': העדפות מוצהרות : בחלק זה המרואיין התבקש לבחון מצבים שונים שקשורים בנהיגה (זמן הנסיעה, אזור הנסיעה, סוג הנסיעה, ליווי בזמן הנסיעה ומזג האוויר בזמן הנסיעה). בכל פעם הוצגו לו 2 חלופות של מצבים והוא היה צריך לבחור ברצוי לו.

מקרב הקשישים שרואיינו: ל-15% רישיון על רכב כבד, אוטובוס או מונית; 69% מהם היו נשואים 20% אלמנים והיתר גרושים (9%) או רווקים (2%). ל-87% מהם יש ילדים כאשר 50% גם גרים בקרבת הילדים שלהם. 5% הצהירו שהם גרים אצל הילדים (המצהירים היו אלמנים או גרושים). 45% מהקשישים הם בעלי השכלה אקדמית ו-32% נוספים בעלי השכלה על תיכונית לא אקדמאית. כרבע מהמרואיינים עובדים, מחציתם כשכירים ומחציתם כעצמאיים. כ-60% מהם עובדים עבודה מלאה ו-20% מהם משתמשים ברכב לצורך נסיעה יומית לעבודה. כ-25% מהקשישים מתנדבים אך רובם (80%) הצהירו

כי ההתנדבות היא ליד הבית ולא דורשת מהם נהיגה. מעניין כי כ-30% מכלל אלה שהצהירו שהם מתנדבים הגדירו את ההתנדבות שלהם כעזרה לילדים או הנכדים.

המאפיינים הדמוגרפיים שהתקבלו עבור קבוצת הסקר תואמים את המאפיינים של אוכלוסיית הקשישים כפי שעולה מנתוני הלמ״ס, מלבד שיעור המועסקים בסקר שהוא גבוה ב-7% נקודות אחוז משיעור המועסקים הקשישים לפי הלמ״ס. תוצאה זו מוסברת בעובדה שהסקר התייחס רק לקשישים נהגים ושניתן להניח כי שיעור המועסקים בקרב קבוצה זו גבוה יותר בשל הצורך שלהם להגיע לעבודה וכן בשל המתאם הקיים בין הפרט העובד ורמת היכולות הקוגניטיביות והפיזיות שלו.

### 4.2 מאפייני הנהיגה העולים מן הסקר

בבדיקת מאפייני הנהיגה של הקשישים שנשאלו בסקר נמצא כי לרוב מאפייניהם תואמים את אלו המתוארים בסקרי הלמ״ס מלבד הבדלים הנובעים מההגדרה המיוחד של המרואיינים. כך לדוגמא שיעור המחזיקים ברכב בפועל בקרב קבוצת הסקר עמד על 83%, כפול מהשיעור המוחזק על ידי הקשישים לפי הלמ״ס, תוצאה הנובעת מכך שנשאלו רק קשישים להם יש רישיון נהיגה וצפוי כי הם אלו שיחזיקו רכב ברשותם. היקף הקשישים בסקר שלהם רישיון נהיגה מיוחד (משאית, מונית ואוטובוס) בנוסף על רישיון על רכב פרטי גבוה יחסית לממוצע כל האוכלוסייה ודומה בשיעוריו לאלה שתוארו לעיל. הקשישים שנסקרו מחזיקים ברישיון שנים רבות – 45 בממוצע אצל הנשים ו-48 בממוצע אצל הגברים. הנסקר עם שנות הרישיון הנמוכות ביותר היה גבר עם 25 שנות ניסיון. רוב הקשישים (65%) נוהגים מדי יום. שיעור זה גבוה במיוחד אצל גברים (84%) ויורד אצל הנשים (46%). גם אלה שלא נוהגים בכל יום נוהגים פעם או פעמיים בשבוע ורק מעטים (5%) אמרו שהם נוהגים בתדירות נמוכה יותר.

בשנת 2016 היה השנתון הממוצע של כלי הרכב המצויים אצל הקשישים – 2010 כלומר בממוצע הקשישים החזיקו רכב בן 5.5 שנים (דיאגרמה 4.1). 3% מהם החזיקו ברכבים הישנים ביותר משנת 1999 (רכבים בני 17 שנה) ו-5% החזיקו ברכבים חדשים משנתון 2016. הגיל הממוצע של כל מצבת כלי הרכב בשנת 2016 עמד על 8.6 שנים (למ״ס), ומכאן שהגיל הממוצע של כלי הרכב אותם מחזיקים הקשישים נמוך בשלוש שנים מזה המאפיין את האוכלוסייה כולה. תוצאה זו משמעותית גם בהינתן העובדה שמרבית כלי הרכב המצויים אצל הקשיש הם בבעלותו (95%) בעוד שבמצבת כלי הרכב הכללית בישראל שיעור כלי הרכב הפרטיים נמוך יותר (86%). רכבי השכרה וליסינג המצויים בשיעור גבוה יותר באוכלוסייה כולה מאופיינים במספר נמוך של שנים על הכביש (רוב כלי הרכב הפרטיים בבעלות שאינה פרטית היו כלי רכב חדשים עד גיל 3 שנים) ואלה היו אמורים להטות את התוצאה לכיוון ההפוך. גם החששות משינוי המאפיינים את אוכלוסיית הקשישים היו אמורים להקשות עליו למכור את הרכב הישן ולקנות חדש ומשום כך להוביל לתוצאה הפוכה. נראה כי הקשישים נותנים חשיבות רבה להחלפת רכב בתוך סל הצריכה שלהם בין היתר בשל החשש להתמודד עם תקלות ובעיות הכרוכות בכלי רכב ישן.

כאשר נשאלו הנסקרים כל כמה זמן אתם מחליפים את המכונית ענו רוב הקשישים ״לפי המצב״ או ״לפי הצורך״, כלומר לא ידעו להגדיר מדיניות ברורה לגבי נושא החלפת כלי הרכב. כ-10% מהנשאלים ענו כי זהו הרכב האחרון שלהם והם לא מצפים להחליפו למרות שלא נמצא מתאם בין תשובה זו ובין גיל הקשיש. כ-30% ענו כי יחליפו מכונית בפרק זמן של פחות מ-6 שנים.

**דיאגרמה 4.1: התפלגות כלי הרכב של הקשישים לפי שנת עלייה על הכביש, נתוני סקר 2016, אחוזים**

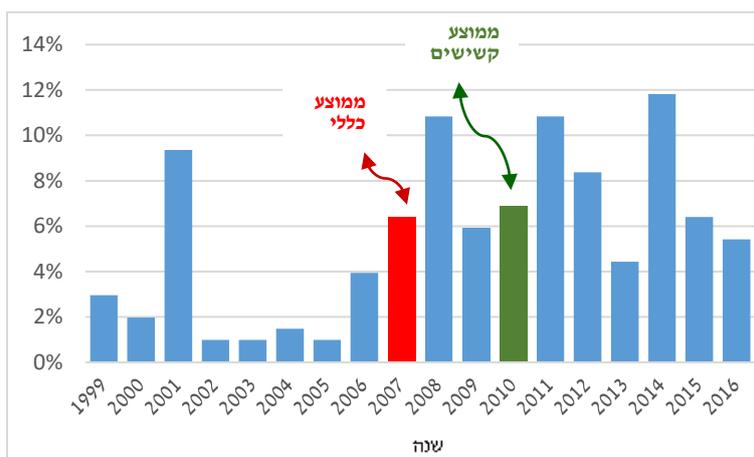

**דיאגרמה 4.2: מועד צפוי להחלפת המכונית, נתוני סקר 2016, אחוזים**

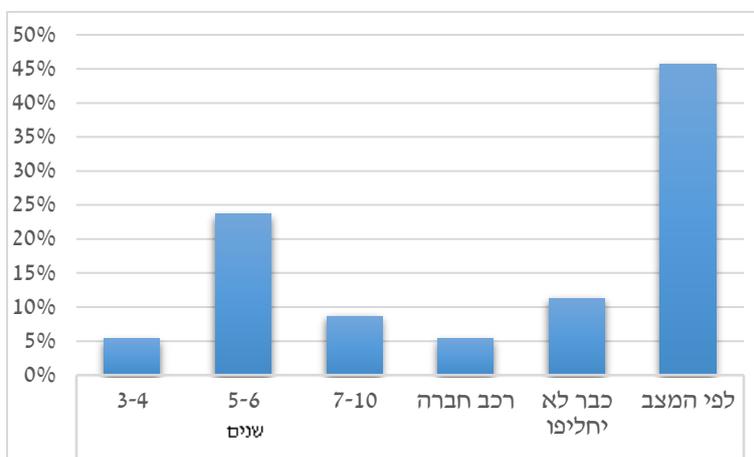

הקשישים נשאלו לגבי הפעולה שיינקטו במידה ותהיה תקלה ברכב. השאלות התייחסו לשלוש תקלות אפשריות: א. תיקון תקר; ב. החלפת מצבר; ג. תקלה אחרת. בכל אחת מהשאלות ניתנה לנשאל אפשרות לבחור בין תיקון באופן עצמאי, הזעקת קרוב משפחה, פנייה לשירות תיקונים או פנייה לגורם אחר (מוסך). 51% מהגברים ו-7% בלבד מהנשים ענו כי הם מחליפים גלגל לבד. במקרה של החלפת מצבר רק 37% מהגברים ו-4% מהנשים ענו כי יעשו זאת באופן עצמאי (דיאגרמה 4.3). לגבי תקלה אחרת כל המשתתפים בסקר ענו כי יפנו לעזרה ולא ינסו לתקן לבד. במקרה שהקשישים מחליטים לא לתקן לבד, נשים יעדיפו לקרוא לבן משפחה בעוד גברים יעדיפו לקרוא לשירות תיקונים. נהגים "כבדים" כלומר כאלה שיש להן רישיון גם על רכב כבד, מונית או אוטובוס, ממשיכים להיות "מחוברים" לרכב גם בגיל המבוגר – 75% ענו שיחליפו גלגל באופן עצמאי ו-85% ענו שיחליפו מצבר לבד. ייתכן כי הפער בתשובות בין תיקון תקר והחלפת מצבר כרוכה ביכולת הפיזית הנדרשת בהחלפת הגלגל ופחות בהחלפת המצבר וזאת למרות הנגישות הנמוכה יותר למצבר.

מגורים ליד ילדים מקטינים את אחוז המתקנים לבד גם אצל הגברים וגם אצל הנשים. כמו כן ירד משמעותית שיעור אלה שאמרו כי יתקנו את התקלה לבד ככל שהנשאל היה מבוגר יותר, במיוחד אצל הגברים: שיעור המתקנים תקר לבד בגיל 70-74 עומד על 58% והוא יורד ל-13% בגילאים 75-80 (דיאגרמה 4.4).

**דיאגרמה 4.3: עזרה בתיקון התקלות ברכב, התפלגות הפניות, נתוני סקר 2016, אחוזים**

4.3A: טיפול בתקר      4.3b: החלפת מצבר

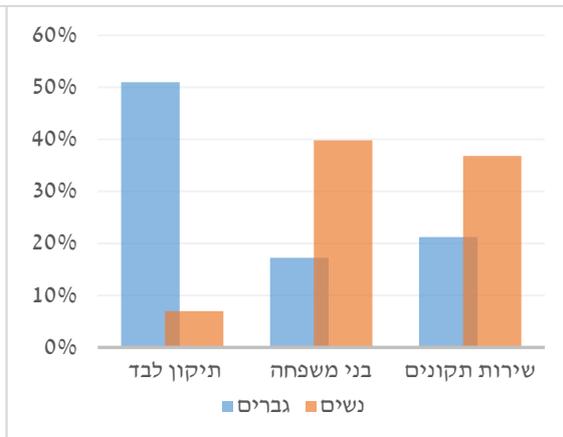
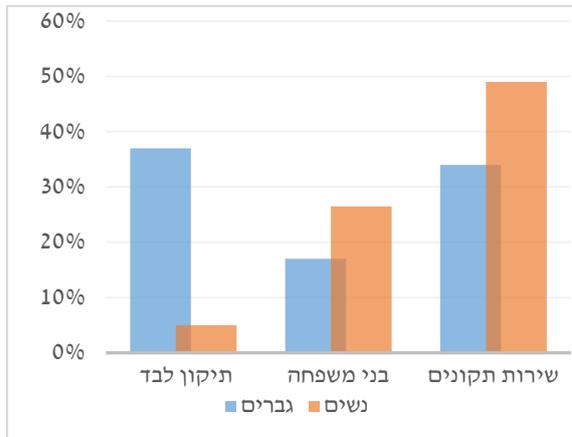

**דיאגרמה 4.4: אחוז הקשישים שבחרו לתקן באופן עצמאי תקר או מצבר, נתוני סקר 2016, אחוזים**

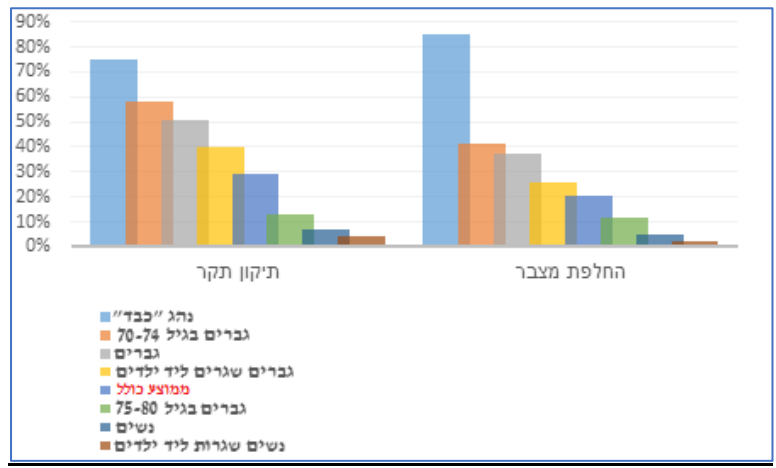

### 4.3 חשיבות מערכות הבטיחות ברכב ותאונות דרכים

במהלך הסקר נשאלו הקשישים לגבי חשיבות 6 מערכות בטיחות לרכב המוצעות היום לציבור. במידה והנשאל לא הכיר את מערכת הבטיחות הוא קיבל הסבר לגביה. עבור כלל מערכת נשאלו הקשישים בשלב ראשון עד כמה הם חושבים שהמערכת חשובה ובשלב השני האם היו מוכנים לשלם כדי להתקין את המערכת במכונית שלהם (מבלי לציין את עלות המערכת).

הקשישים רואים במערכות הבטיחות כלי חשוב (לוח 4.1). בודדים ענו שמערכות אלו לא חשובות ורוב הקשישים ענו כי מערכות הביטחון חשובות או חשובות מאוד. החשיבות הגבוהה ביותר ניתנה למערכת למניעת החלקה (ABS) ומצלמת הנסיעה לאחור (רוורס) לרכב. ראוי לציין כי פער ביחס בין גברים לנשים למערכות בטיחות נמצא מובהק במערכות לשמירת מרחק וסטייה ומצלמת הנסיעה לאחור לרכב. גברים ייחסו באופן מובהק חשיבות גדולה יותר לשני המשתנים הללו, ובמיוחד למצלמת הנסיעה לאחור. בדיקה נוספת הראתה כי אין הבדלים מובהקים בין קשישים בעלי השכלה תיכונית ובין בעלי השכלה על תיכונית ביחס למערכות הבטיחות מלבד ביחס למערכת שמירת מרחק וסטייה מהנתיב. עבור מערכת זו בעלי

השכלה על תיכוניות היו יותר נחרצים בעמדותיהם – אחוז גבוה מהם (יחסית לקבוצה השנייה) טען שמערכת זו חשובה מאוד או לא חשובה בכלל.

למרות החשיבות הרבה שהקשישים מקנים למערכות הבטיחות ברכב הם ענו באופן גורף בשלילה לשאלה האם יסכימו להתקין מערכות אלו בתשלום (97% מהנשאלים). פערים דומים בתשובות יתוארו גם בהמשך המחקר. ייתכן כי הקשישים קושרים בין התקנת מערכות הבטיחות לבין נזקקות או חולשה של הנשאל בהקשרים הקשורים לנהיגה שלו. מאחר והנהיגה היא חלק מתחושת הביטחון והעצמאות של הקשיש, כל ניסיון לרמוז כי יש צורך בחיזוק בנושא, למשל על ידי התקנת מערכות בטיחות, נדחים על הקשישים.

**לוח 4.1: חשיבות מערכות הבטיחות ברכב בעיני המרואיינים, נתוני הסקר 2016**

| לא כ"כ חשוב | חשוב | חשוב מאוד | |
|---|---|---|---|
| 3.9% | 60.6% | 39.4% | מערכת למניעת אובדן השליטה על הרכב |
| 1% | 21.9% | 77.1% | מערכת למניעת החלקה ABS |
| 8.2% | 63.4% | 28.3% | מערכת לבדיקת לחץ אוויר בגלגלים |
| 8.8% | 67.2% | 24.0% | מערכת לשמירת מרחק ומניעת סטייה מהנתיב (AWACS) |
| 4.4% | 24.4% | 71.2% | אביזרי הבטיחות מצלמת רוורס |

בשלב האחרון של חלק זה נשאלו הקשישים לגבי תאונות דרכים שעברו. השאלה הראשונה שעלה האם עברו מתי שהוא בחייהם תאונת דרכים ואם כן, מתי. רק 12.2% מהנשאלים (25 איש מתוך 205) ענו שעברו בחייהם תאונת דרכים, כאשר רובם ככולם (למעט שני נשאלים) עברו את תאונת הדרכים ב-5 שנים האחרונות. 92% דיווחו על התאונה כקלה. המספר הנמוך של הנשאלים שענו שעברו תאונת דרכים מפתיע מאוד בהינתן שהסיכוי להיפגע בתאונת דרכים במהלך החיים ועד גיל 70 גבוה פי שלושה[5]. נראה כי מדובר בהטיה של הקשיש לא לזכור תאונות ובמיוחד לא את אלה שהיו לפני זמן. ההטיה לכיוון התאונות הקלות יכולה להיות מוסברת בעובדה שמי שנפגע קשה (או נהרג בתאונת הדרכים), במיוחד בחמש השנים האחרונות, הסיכוי כי ימצא לבדו במקומות בהם נבחרו הנסקרים קטן.

## 5. הקשיים בנהיגת הקשיש והפרישה מנהיגה

העדפות הנשאלים הקשישים לגבי יכולות הנהיגה שלהם, הקשיים בנהיגה וההחלטה לפרוש מנהיגה נבחנו מכמה זוויות. ראשית נשאלו הנסקרים לגבי היחס של הילדים או בני המשפחה בהקשר לנהיגתו של הנשאל. לאחר מכן הנסקרים נשאלו לגבי דעתם בנוגע למגבלות נהיגה שונות המאפיינות את הגיל. בשלב השלישי נשאל הקשיש באופן קונקרטי האם הוא אישית חש במגבלות כלשהן וכן האם יש לו חששות מנהיגה, ובשלב האחרון של חלק זה נשאלו הקשישים לגבי רמת הסכמתם לשורה

---

[5] הערכה ראשוני לסיכוי להיפגע בתאונות דרכים ניתן לעשות על ידי לקיחת שיעור הנפגעים בשנה, חלוקתו במספר הנהגים וחישוב ההסתברות המצטברת ל-50 שנה (מגיל 20 עד גיל 70). אם ניקח את ההסתברות הממוצעת של נהג לעבור תאונת דרכים בעשור האחרון, 0.7% בשנה, חישוב זה מביא אותנו ל-35% - סיכוי מצטבר של נהג קשיש להיפגע בתאונת דרכים במהלך החיים. ראוי לציין כי הסיכוי ירד בעשור האחרון עם ירידת שיעור התאונות כלומר אומדן זה מוטה כלפי מטה.

של היגדים הקשורים לנהיגת קשישים כאשר ההיגדים התייחסו לקשיים פיזיים ומנטאליים של נהיגת קשישים. סה"כ נשאלו בחלק זה 25 שאלות אשר תקפו את הנושא מנקודות מבט שונות.

כשליש מהקשישים שנדגמו קיבלו הצעה כלשהיא מבני משפחתם הנוגעת לנהיגה שלהם. שיעור זה גבוה יותר אצל הנשים (45%) ועולה משמעותית ל-79% אצל קבוצת הגיל המבוגרת יותר (גילאי 75-80). מבין האפשרויות שהוצגו למרואיינים התקבל כי שיעור הקשישים שקיבלו הצעה "לתלות את המפתחות" ולהפסיק לנהוג הוא הגבוה ביותר (10%), ואחריהם הצעה לבדיקת כשירות הנהיגה (8%), צמצום הנהיגה בלילה (7%) והצעה לשימוש בתחבורה הציבורית (7%). לא נמצא מתאם בין אלו שקיבלו הצעות מילדיהם לגבי הנהיגה ובין המרואיינים שהצהירו כי יפסיקו את הנהיגה שלהם בהתאם למצב או שהם חוששים מנהיגה.

בשלב השני נשאלו הקשישים ביחס מגבלות שונות המאפיינות את אוכלוסיית הקשישים ויכולות להגביל את הנהיגה שלהם. כמחצית מהמרואיינים סברו שיש להגביל את הנהיגה במקרה של בעיות ראיה או הזזת הגוף ו-30% סברו שיש להגביל את הנהיגה במקרה של בעיות שמיעה או מחלת לב. עם זאת, כאשר נשאלו המרואיינים שאלות כללית לגבי נהיגת הקשישים מסוג : "האם יש להגביל אנשים מבוגרים בנהיגה" או "האם יש להתנות את חידוש הרישיון במבחן כלשהו?" השיבו כמעט כל הנשאלים (94%) תשובה שלילית לשתי השאלות. המרואיינים שענו "כן" לשאלה לגבי הגבלת הסיעה הוסיפו כי יש להגביל את הנהיגה רק כפונקציה של מצב הבריאות. אלה שענו כן לשאלה "לגבי התניה במבחן כלשהו הוסיפו כי "המבחנים כל 5 שנים הקיימים כיום מספיקים". נראה כי קיים פער בין ההבנה שמגבלות פיסיות או מחלות צפויות להשפיע על איכות הנהיגה ובין הקישור של מגבלות אלה לאוכלוסיית הקשישים באופן כללי.

המרואיינים היו מאוד בטוחים בנהיגה שלהם. 85% מהם אמרו שהם לא מתכננים להפסיק לנהוג. אלה המתכננים להפסיק לנהוג לא תלו את הפסקת הנהיגה בגיל אלא טענו כי יפסקו כאשר יתדרדר מצב בריאותם או ההרגשה הכללית שלהם. כמעט כל המרואיינים (למעט חמישה) טענו כי אין להם חששות מהנסיעה. החוששים כולם חששו מהנהיגה בלילה. בהינתן כי שליש מהנשאלים קיבלו הערות מילדיהם לגבי מצב הנהיגה שלהם נראה כי הנשאלים בחרו להתעלם מהצעות הילדים ונשארו עקביים בתשובותיהם שאין צורך להטיל עליהם מגבלות כלשהן.

התוצאות מצביעות כי הנשאלים אינם רואים בגיל מגבלה כלשהי. התשובות השליליות חזקות בעוצמתן בכל ניסיון להגביל את הנהיגה מסיבות הקשורות בגיל. למרות הקשר שיש בין רמת הקשיים הפיזיים והמנטליים והגיל המבוגר - כך לדוגמא וכפי שראינו, החיוב להרכיב משקפיים עלה משמעותית בקבוצת הגיל שמעל 65 – המרואיינים לא קושרים בין הדברים הפסקת הנהיגה קשורה אצלם רק למגבלות פיזיות ולא לגיל. יתר על כן, גם מגבלות שהיינו מצפים בהם לתשובה חיובית חזקה יותר כגון : "האם יש להגביל נסיעה בשל מגבלת ראיה" מחצית מהנשאלים השיבו בשלילה וחשבו כי אין להגביל את הנסיעה גם במקרה זה. נראה כי החשש בפני אובדן יכולת הנהיגה מונע מהקשישים לענות חיובית על אפשרות של הטלת מגבלות במקרים אלו וכן הם נמנעים מכל הכללה של המגבלות על אוכלוסיית הקשישים, גם אם היא נכונה בממוצע.

ולסיום, הנשאלים התבקשו להתייחס לתשעה היגדים הקשורים לנהיגת קשישים, בכל היגד התבקשו הנשאלים לדרג את רמת הסכמתם להיגד בסולם מדרג של 1 עד 6 (לוח 5.1).

ההיגדים נוסחו כך שחלקם ייחסו "לאנשים מבוגרים" באופן כללי וחלקם לנשאל עצמו וכן יש בהם ערבוב של היגדים הנוגעים למגבלות פיזיות של אנשים מבוגרים (קושי בנהיגה לאחור, שימוש מופרז בבלמים, קושי בחנייה) ולמגבלות מנטליות (כועסים ומתרגזים יותר בזמן חניה, מתקשים לזהות תמרורים וסימנים, מגיבים פחות טוב למצבים בלתי צפויים). שני ההיגדים התייחסו לנהיגת קשישים באופן כללי וטענו כי הם שולטים פחות טוב על הרכב בדרך כלל וכן שהם לא מודעים תמיד למגבלות שלהם בנהיגה. היגד אחד התייחס ליחס הסביבה וטען כי הנהגים האחרים "מרבים לצפור" לנהגים מבוגרים.

**לוח 5.1: התפלגות רמת ההשכלה להיגדים שונים הקשורים לנהיגת "אנשים מבוגרים", סקר מיוחד, 2016**

| סיווג | | ההיגד | לא מסכים כלל | במידה מעטה מאד | במידה מעטה | במידה מסוימת | במידה רבה | במידה רבה מאד |
|---|---|---|---|---|---|---|---|---|
| קשיים טכניים בנהיגה | 1. | לאנשים מבוגרים יש קושי בנהיגה לאחור (רוורס) | 2.96% | 5.42% | 4.93% | 39.90% | 25.12% | 21.67% |
| | 2. | אנשים מבוגרים משתמשים באופן מופרז בדוושת הבלמים | 3.47% | 6.93% | 11.39% | 39.60% | 34.16% | 4.46% |
| | 3. | אנשים מבוגרים מתקשים יותר בחניה | 2.0% | 3.6% | 3.6% | 37.1% | 29.4% | 24.4% |
| קשיים מנטליים בנהיגה | 4. | אנשים מבוגרים מגיבים פחות טוב למצבים בלתי צפויים בדרך | 13.0% | 21.0% | 16.5% | 43.5% | 5.0% | 1.0% |
| | 5. | אנשים מבוגרים כועסים ומתרגזים יותר בזמן הנהיגה | 8.5% | 9.5% | 28.4% | 39.8% | 12.4% | 1.5% |
| | 6. | אנשים מבוגרים מתקשים לזהות תמרורים וסימנים בצדי הדרך | 13.4% | 8.9% | 21.3% | 43.6% | 10.9% | 2.0% |
| הערכת איכות הנהיגה ומודעות | 7. | אנשים מבוגרים שולטים פחות טוב על הרכב בדרך כלל | 59.8% | 21.6% | 5.5% | 11.1% | 2.0% | |
| | 8. | אנשים מבוגרים לא מודעים תמיד למגבלות שלהם בקשר לנהיגה | 5.6% | 18.9% | 18.4% | 51.0% | 5.1% | 1.0% |
| יחס הסביבה | 9. | נהגים אחרים מרבים לצפור לאנשים מבוגרים | 1.5% | 8.4% | 26.1% | 35.0% | 25.6% | 3.4% |

רוב הקשישים חושבים שיש קשיים טכניים הכרוכים בנהיגת קשישים. בממוצע 85% אמרו שהם מסכימים במידה מסוימת עד הסכמה במידה רבה לעובדה שקשישים מתקשים בנהיגה לאחור, משתמשים בבלמים באופן מופרז ומתקשים בחניה, מתוכם 45% בממוצע אמרו שהם מסכימים בתחומים אלו במידה רבה או במידה רבה מאוד. גם שיעור ההסכמה לגבי ההיגדים המתייחסים לקשיים מנטליים לא נמוך: 50% בממוצע הסכימו במידה מסוימת עד במידה רבה מאוד עם ההיגדים המתייחסים ליכולת תגובה של הקשיש, כעס בזמן נסיעה או יכולת זיהוי. שיעור זה יורד משמעותית ל-11% כאשר מסתכלים על אלו שמסכימים עם ההיגדים במידה רבה או במידה רבה מאוד (דיאגרמה 5.1).

למרות ההסכמה הרחבה בהיגדים הספציפיים לגבי הקשיים בנהיגת קשישים, כאשר נשאלו המרואיינים באופן ישיר אם הם מסכימים ש"אנשים מבוגרים שולטים פחות טוב על הרכב בדרך כלל" 60% טענו שהם לא מסכימים בכלל עם היגד זה. רק 13% הסכימו במידה מסוימת או במידה רבה עם היגד זה ואף אחד מהנשאלים לא הסכים עם ההיגד במידה רבה מאוד.

הדבר מעיד על פער ביכולת ההבחנה או ההשלמה של הקשיש בקושי מוגדר, טכני או מנטלי, ובין יכולת ההפנמה של הקשיים הללו כמשפיעים על הנהיגה כמכלול. השתקפות יפה של בעיה זו ניתן לראות במידת ההזדהות של הנשאלים להיגד : "אנשים מבוגרים לא מודעים תמיד למגבלות שלהם בקשר לנהיגה". להיגד זה הסכימו 67% מהנשאלים (51% "במידה מסוימת") ורק 6% לא הסכימו בכלל.

מרבית הקשישים הסכימו כי נהגים אחרים מרבים לצפור לנהגים קשישים (64% הסכימו עם ההיגד). עם זאת לא נמצא מתאם גבוה בין התשובה לשאלה זו המצביעה על כך שנהגים אחרים (שצופרים) מבחינים בקושי מסוים בנהיגת הקשיש ובין התשובה על השאלה "אנשים מבוגרים שולטים פחות טוב על הרכב בדרך כלל". כלומר, למרות ההבחנה בצפירות מסביב, הקשיש אינו תולה את הסיבה לצפירות בנהיגתו שלו אלא בהתנהגות בעייתית של הצופר.

הפערים שנמצאו בתשובות הקשישים בתשובות על ההיגדים מציגים פערים דומים לאלו שעלו בסעיפים הקודמים. המרואיינים במקרה זה מודעים לכך שמגבלות טכניות או מנטליות, שהן מגבלות המאפיינות קשישים, פוגעות באיכות הנהיגה. הם מסכימים שיש פעולות במסגרת הנהיגה שהקשיש מתקשה איתם כאשר ההסכמה היא יותר רחבה במקרה של קשיים פיזיים ופחות רחבה בקשיים מנטליים-קוגניטיביים. עם זאת, הם מסרבים להכיר במשמעות התשובות שענו ביחס לכלל אוכלוסיית הקשישים וכן ביחס לעצמם כחלק מאוכלוסייה זו. הם גם אינם מסכימים שנהג קשיש שולט על הרכב פחות טוב, לא באופן כללי ולא ביחס לעצמם. עם זאת, הם מודים כי ייתכן שקשישים אינם מודעים למגבלות שלהם.

**דיאגרמה 5.1: שיעור המסכימים עם ההיגדים בחלוקה לקשיים טכניים וקשיים מנטליים בנהיגה, סקר מיוחד 2016, אחוזים**

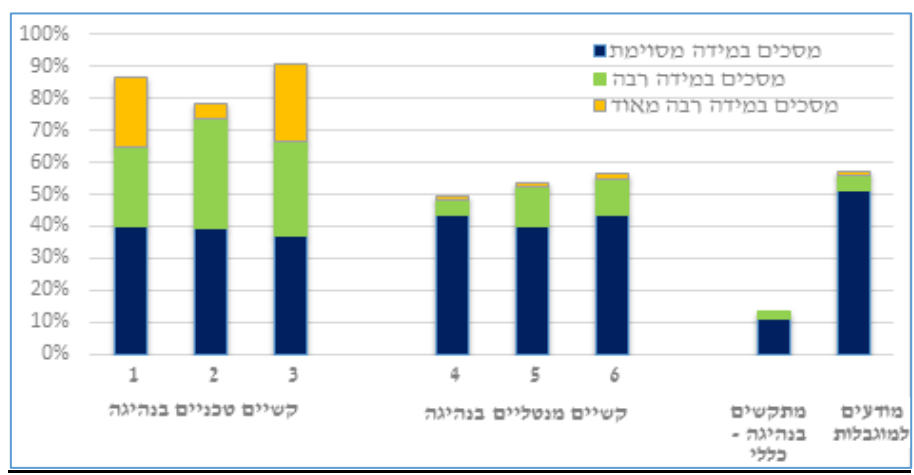

## 6. בחינת העדפות וחסמי נסיעה של הנהגים הקשישים

### 6.1 מתודולוגיה והאפיון הסטטיסטי של שיטת העדפות המוצהרות

במהלך הריאיון בוצע סקר בשיטת ההעדפות המוצהרות לבחינת העדפות וחסמי הנסיעה של הנהגים הקשישים (חלק 6 בשאלות הסקר). סקר העדפות מוצהרות הינו שיטת איסוף נתונים באמצעות הצגת מצבים שונים המאופיינים בכמה פרמטרים, כאשר המרואיין מתבקש לבחור את האפשרות העדיפה עליו או להחליט שאין הבדל בין האפשרויות העומדות בפניו. כל אלטרנטיבה מציגה מקבץ שונה של מאפייני הנסיעה, לדוגמא : נסיעה עירונית, באור יום, למרחק קצר, עם נוסעים

ובמזג אויר גרוע, לעומת נסיעה עירונית, בלילה, למרחק קצר, לבד ובמזג אויר טוב. בצורה זו ניתן לקבל את החשיבות של כל מאפיין ומאפיין וכן לאמוד את רמת התועלת המתקבלת למרואיין מרמה שונה של כל מאפיין.

סקר העדפות מוצהרות בחירה בין שני תרחישים המוצגים למרואיין שווה לתצפית אחת. ההנחה היא שהתצפיות למרואיין בלתי תלויות. לפיכך, אם מתקבלים מכל מרואיין 5 בחירות (תצפיות), מדגם בגודל 200 מרואיינים (כפי שבוצע בסקר זה) ייתן 1000 תצפיות שאמורות לבנות מודל שיהיה מובהק. גודל המדגם אפשר בניית מודלים מובהקים גם לתת-אוכלוסיות בקרב אוכלוסיית הנהגים הקשישים.

ההנחה הבסיסית היא שכאשר אנשים מחליטים לבחור חלופה מסוימת הם מעניקים לה תועלת גבוהה יותר מאחר החלופה האחרת. התועלת שאנשים מעניקים לחלופה היא בחלקה מבוססת על משתנים הניתנים למדידה ובחלקה על גורמים שאינם מדידים ומוגדרים כשארית ($e$). מודל ה-$LOGIT$ שמשתמשים בו מניח התפלגות מערכית של השגיאה.

פונקציית התועלת המוגדרת בעבודה זו בחנה את השפעתם של חמישה משתנים:

6.1  $U = a_1 \times Hour + a_2 \times Area + a_3 \times Disrance + a_4 \times partners + a_5 \times weather + e$

כאשר, $a_1, a_2, a_3, \ldots, a_n$ הם המקדמים שנבנו על ידי המודל, ומהווים למעשה "משקלות העדפה" (Utilities) למאפיינים השונים המוגדרים בלוח 6.1.

**לוח 6.1: המשתנים המגדירים את סקר ההעדפות ורמותיהם**

| מאפיין | רמות המאפיין |
|---|---|
| שעת הנסיעה | נסיעה באור יום |
|  | נסיעה בשעות הלילה |
| אזור הנסיעה | נסיעה עירונית |
|  | נסיעה בינעירונית |
| מרחק הנסיעה | מרחק נסיעה ארוך |
|  | מרחק נסיעה קצר |
| נסיעה עם נוסעים נוספים | נסיעה לבד |
|  | נסיעה עם נוסעים נוספים |
| מזג אוויר בזמן הנסיעה | נסיעה במזג אוויר נח |
|  | נסיעה במזג אוויר חורפי |

המודל מוצא מקסימום להסתברות הבחירה של פונקציה לוגריתמית מהצורה:

6.2  $P = \dfrac{e^U}{1 + e^U}$

תוצאות המחקר הן המשקלות שניתנו לכל אחד מהמאפיינים שהוגדרו, המגדילים את התועלת היחסית שיש לפרט מאפיין זה, והם מציגות את ההעדפה היחסית בין התכונות השונות והחוזק של כל תכונה.

בחינת מובהקות המודל מתבצעת על ידי שני מדדים: א. מדד ה-Chi-square מדד זה נבנה כפעמיים ההפרש בין Loglikelihood ללא מודל ובין ה-Loglikelihood של המודל. במקרה שלנו כל מדד Chi-square עם 5 דרגות חופש (מספר הפרמטרים שנמדדו במודל) שערכו מעל ל-15 הינו מובהק ברמת מובהקות של 99%. ב. מדד ה- Rho-Square מדד זה

הינו מדד שמדמה את מדד ה R-Square ברגרסיה אבל ערכיו נמוכים. ערך של Rho-Square של 0.2 מקביל בערכו ל- R-Square של 0.7.

המודל הורץ על כל הנסקרים יחד וכן בתת קבוצות בסיווג לפי מאפיינים דמוגרפים. תתי הקבוצות שנבנו אפשרו את ההשוואות הבאות: א. נשים מול גברים; ב. גילאי 70-74 מול גילאי 75-80; ג. נוסעים בתדירות גבוהה מול אלה הנוסעים בתדירות נמוכה; ד. נשואים מול לא נשואים; ה. קשישים הגרים ליד ילדיהם לעומת אלה שלא; ו. קשישים בעלי רמות השכלה שונות; ז. קשישים עובדים מול אלה שלא עובדים. כל ההרצות – ברמה הכללית ובתתי הקבוצות - יצאו מובהקים כפי שמפורט בלוח 6.2 (פירוט התוצאות המלא של המודל עבור תתי הקבוצות מופיע בנספח 2).

**לוח 6.2: מודל העדפות - מדדים סטטיסטים של כל המודלים שהורצו בהתאם לאפיון הדמוגרפי, נתוני סקר מיוחד 2016**

| Rho-square: | Chi Square | Final log-likelihood: | Init log-likelihood: | מספר הנשאלים | |
|---|---|---|---|---|---|
| 0.2244 | 705.7 | -750.1 | -1103.0 | 201 | כל המרואיינים |
| 0.2317 | 481.5 | -478.9 | -719.6 | 131 | נוסעים כל יום או כמעט כל יום |
| 0.2265 | 238.6 | -248.8 | -368.0 | 67 | נוסעים בתדירות נמוכה יחסית |
| 0.2299 | 577.9 | -584.4 | -873.4 | 159 | בגיל 70 עד 74 |
| 0.2166 | 139.5 | -159.9 | -229.6 | 42 | בגיל 75 ומעלה |
| 0.2198 | 334.4 | -371.1 | -538.3 | 98 | גברים |
| 0.2351 | 374.0 | -361.2 | -548.2 | 100 | נשים |
| 0.1971 | 137.6 | -194.9 | -263.7 | 48 | השכלה עד תיכון כולל |
| 0.2499 | 347.7 | -297.5 | -471.3 | 86 | השכלה על תיכונית |
| 0.2559 | 271.1 | -221.5 | -357.0 | 65 | השכלה אקדמאית |
| 0.2067 | 139.7 | -177.3 | -247.2 | 45 | עובד/ת |
| 0.2402 | 600.7 | -555.5 | -855.8 | 156 | לא עובד/ת |
| 0.2167 | 467.7 | -535.2 | -769.0 | 140 | נשואים |
| 0.2606 | 259.1 | -204.4 | -334.0 | 61 | לא נשואים |
| 0.2327 | 373.4 | -368.1 | -554.8 | 101 | גרים ליד הילדים |
| 0.2221 | 317.9 | -345.3 | -504.3 | 92 | לא גרים ליד הילדים |

המקדמים הנאמדים של המאפיינים (המפורטים בלוח 6.3) משקפים את עוצמת ההשפעה של כל מאפיין נסיעה. מקדם של מאפיין שלילי פרושו שהתועלת של המרואיין ממאפיין זה היא שלילית. ההפרש בין המקדם הגבוה של רמת מאפיין מסוים לבין הערך הנמוך של רמת אותו מאפיין מציין את החשיבות של המאפיין, לדוגמא אם מקדם לנסיעה באור יום הוא 0.9593 ומקדם לנסיעה בשעות הלילה הוא -0.9593 (נסיעה בשעות הלילה הינה שלילית למרואיין) אזי החשיבות של המאפיין היא 1.9186. החשיבות היחסית מחושבת כחלקה של החשיבות של מאפיין מסוים מסך החשיבויות של כל המאפיינים.

לצורך השוואה בין מודלים שונים אנו משתמשים בשני פרמטרים: א. השוואה בין החשיבויות היחסיות של המאפיינים השונים כאשר השוואה זו אינה סטטיסטית ולכן לא ניתן לציין אם היא מובהקת. ב. השוואה בין המקדמים השונים של

המאפיינים בשתי ההרצות. ההשוואה מתבצעת באמצעות מבחן t-test המאפשר זיהוי של מובהקות ההפרש בין המקדמים במודלים השונים.

### 6.2 ממצאי מודל העדפות לכלל המרואיינים

בסה"כ נתקבלו תוצאות כל 5 בחירות של 205 מרואיינים (סה"כ 1020 תרחישים). ב-36% מהתרחישים המרואיינים החליטו לא לבחור באף אחת מהחלופות שהוצגו בפניהם כלומר היו אדישים בין שתי האפשרויות. סיכום תוצאות המודל לכל המרואיינים מפורט בלוח 6.3.

**לוח 6.3: מודל העדפות לכלל המרואיינים – תוצאות ההרצה**

| המאפיין | רמת המאפיין | מקדם | שגיאת תקן | T-TEST | חשיבות יחסית |
|---|---|---|---|---|---|
| **שעת הנסיעה** | נסיעה באור יום | 0.9593 | 0.07069 | 13.5696 | **30.18%** |
| | נסיעה בלילה | -0.9593 | 0.07069 | -13.5696 | |
| **אזור הנסיעה** | נסיעה עירונית | 0.11004 | 0.05781 | 1.9035 | **3.46%** |
| | נסיעה בינעירונית | -0.11004 | 0.05781 | -1.9035 | |
| **מרחק הנסיעה** | מרחק נסיעה ארוך | -0.77955 | 0.067 | -11.6344 | **24.53%** |
| | מרחק נסיעה קצר | 0.77955 | 0.067 | 11.63436 | |
| **נוסעים ברכב** | נסיעה לבד | 0.05139 | 0.05771 | 0.89047 | **1.62%** |
| | נסיעה עם מישהו נוסף | -0.05139 | 0.05771 | -0.89047 | |
| **מזג אוויר בנסיעה** | נסיעה במזג אוויר נח | 1.27825 | 0.07858 | 16.26668 | **40.22%** |
| | נסיעה במזג אוויר חורפי | -1.27825 | 0.07858 | -16.2667 | |

אם ניקח את חמשת הפרמטרים יחד ונסווג אותם בהתאם לחשיבות היחסית שלהם (כאשר כולם יחד מרכיבים 100%) נמצא כי הקשישים מייחסים את החשיבות הגבוהה ביותר למזג אוויר נח: 40% מכלל התרומה של המאפיינים השונים. הפרמטר השני בחשיבותו הוא נסיעה ביום ולא בשעות הלילה (30%) ומרחק נסיעה קצר (25%). נסיעה עם מלווה והסיווג בין נסיעה עירונית לנסיעה בין עירונית אינם חשובים בעיני הקשישים. ההתפלגות הלא שווה בין המאפיינים מצביעה כי יש כאלה המקשים יותר על נהיגת הקשישים ולכן הם נתונים להם חשיבות גבוהה יותר, כלומר על אף העובדה שהקשישים ענו שהם לא חוששים כלל לנהיגה שלהם אנו מוצעים שהם כן חוששים מנהיגה בתנאי דרך שהם לא אופטימליים.

### 6.3 השוואת החשיבות היחסית של המשתנים בין הקבוצות השונות

על מנת לבחון את החשיבות היחסית שמעניקות קבוצות שונות באוכלוסייה לפרמטרים נבחנה כל קבוצה בנפרד ונעשו השוואות בין הקבוצות השונות ומבחנים סטטיסטים להבדל התוצאות כפי שפורט לעיל (תוצאות ההרצה לקבוצות השונות מוצגות בדיאגרמה 6.1. ההרצות לכל קבוצה בנפרד וההשוואה ביניהן מוצגות בנספח 2)
נשים מייחסות חשיבות רבה יותר מגברים לנסיעה בשעות היום לעומת נסיעה בשעות הלילה – 32.5% לעומת 26.8%. הבדל זה נמצא מובהק. ייתכן כי הדבר קשור לנושא הביטחון בנהיגה אשר מאפיין יותר גברים והתחזקות חוסר הביטחון אצל נשים בשעות הלילה. עם זאת, ייתכן גם שהדבר לא קשור להרגשת הביטחון אלא לגורמים פיזיולוגים כגון התדרדרות יכולת הראיה

אצל נשים מעל גיל 70 שעלולה להיות חזקה יותר אצל נשים ומשום כך פוגעת גם ביכולת הנהיגה בלילה ומחזקת את חשיבות משתנה זה.

נמצא כי גם האוכלוסייה הקשישה יותר (80-75) מקנה חשיבות גבוהה יותר מהצעירים יחסית (גילאי 74-70) למזג אוויר בנסיעה וכן לנסיעה בשעות היום. לעומת זאת מרחק הנסיעה פחות מטריד אותם. כלומר שתי החולשות שחשיבותם היא הגבוהה ביותר מתעצמות מתחזקת עם עליית הגיל. עם זאת, הבדלים אלה בין הקבוצות לא נמצאו מובהקים.

בהשוואה בין הנהגים שנוסעים בתדירות גבוהה (כל יום) לבין אלו הנוסעים בתדירות נמוכה יותר נמצא כי באופן מובהק כי מרחק הנסיעה הינו בעל חשיבות פחותה אצל הנוסעים בתדירות נמוכה ולעומת זאת נסיעה באור יום הינה בעלת חשיבות גבוהה יותר בקרב נוסעים אלו. כמו כן נמצא כי לנוסעים בתדירות נמוכה חשוב הימצאות מלווה בשיעור גבוה יותר לעומת אלו הנוסעים בתדירות גבוהה. הממצאים מתאימים לצפוי ולהנחה כי אלה הנוסעים בתדירות גבוהה הם בעלי מיומנויות נהיגה גבוהות מאלו הנוסעים בתדירות נמוכה. משום כך, הנסיעה בשעות היום והעובדה שהם נוסעים עם מלווה ברכב פחות חשובה עבורם. לעומת זאת, מאחר והם נוסעים כמעט כל יום חשוב להם לא ל"בזבז" זמן רב בכבישים ומשום כך מרחק הנסיעה יותר חשוב עבורם באופן מובהק.

השוואת החשיבות היחסית בהתאם למצב המשפחתי של הנהג העלתה את הממצאים הבאים: שעת הנסיעה ומזג האוויר בנסיעה חשובים יותר לאלו שאינם נשואים באופן מובהק. לעומת זאת מרחק הנסיעה חשוב להם פחות (אך לא באופן מובהק). העדפות אלה שאינם נשואים דומות להעדפות הנוסעים בתדירות נמוכה, ואכן תדירות הנסיעה של הלא נשואים נמוכה מזו של הנשואים.

השוואה נעשתה גם בהתאם למרחק המגורים של הקשיש מבני משפחתו. ההנחה הייתה כי ככל שהקשיש יגור ליד בני משפחתו הוא ימעט יותר בנסיעות ולכן הקשיים הכרוכים במזג האוויר ובנהיגה בשעות הלילה יתעצמו. למרות זאת, לא נמצא הבדל מובהק בקשיים בנהיגה בין שתי הקבוצות ונראה כי בניגוד לתוצאות שהצגנו לעיל (סעיף 4) שהראו כי הקשיש נעזר יותר בבני משפחתו אם הוא גר לידם, במקרה של העדפות הנסיעה לא נמצאה לקרבה המשפחתית השפעה כלשהי.

השוואה נעשתה גם בין הנהגים בהתאם לרמת ההשכלה שלהם. הקבוצה חולקה לשלוש תתי-קבוצות - בעלי השכלה עד תיכונית, השכלה בין 12 ל-15 שנות לימוד ובעלי השכלה אקדמאית. נמצא כי בעלי השכלה עד תיכונית מייחסים חשיבות מועטה יחסית לנסיעה באור יום וחשיבות גבוהה יותר ליתר המגבלות יחסית לבעלי השכלה על תיכונית ואקדמאית. הבדלים ביחס לנהיגה ביום ובמזג אוויר טוב נמצאו מובהקים. בהשוואה בין קשישים עובדים לל עובדים נמצא הבדל מובהק ברמת החשיבות של שעת הנסיעה: קשישים לא עובדים מקנים לנסיעה בשעות יום חשיבות גבוהה יותר לעומת קשישים שאינם עובדים.

**דיאגרמה 6.1: השוואת החשיבות היחסית של מאפייני הנהיגה של קבוצות דמוגרפיות שונות, סקר 2016, אחוזים**

א. השוואה לפי מגדר:

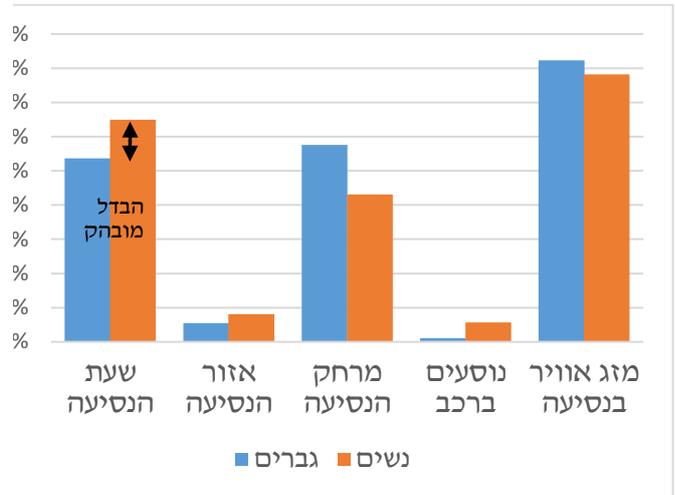

ד. השוואה לפי מצב משפחתי:

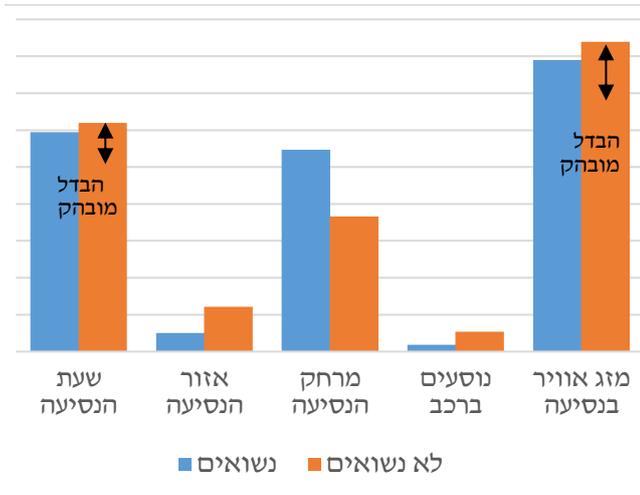

ב. השוואה לפי השכלה:

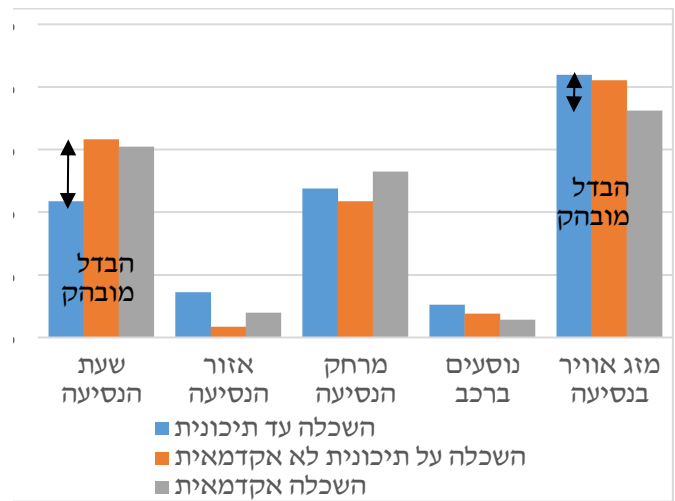

ה. השוואה לפי מגורים ליד ילדים:

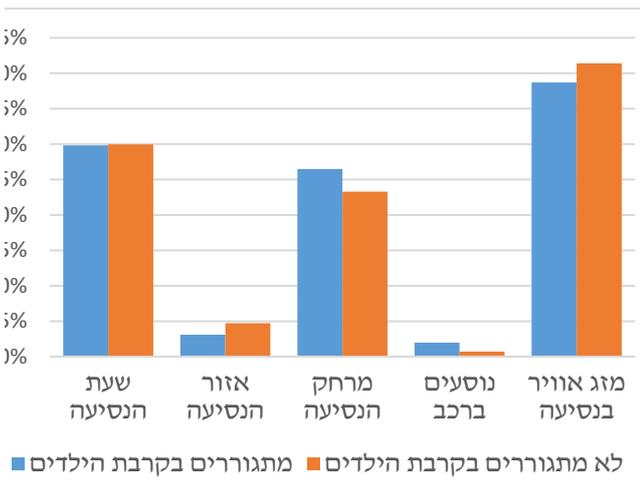

ג. השוואה לפי תדירות נסיעה:

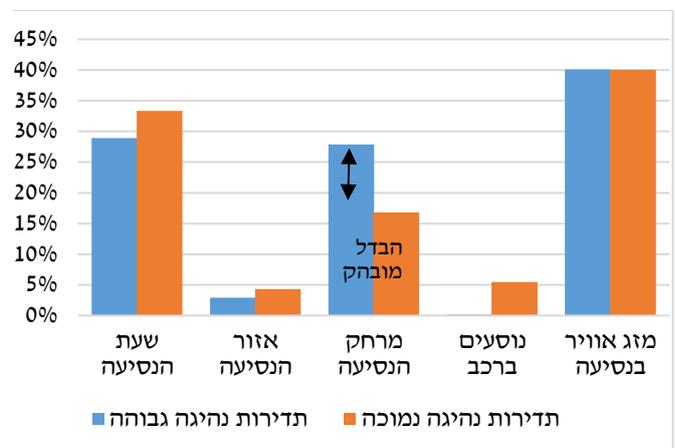

ו. השוואה לפי מצב תעסוקה:

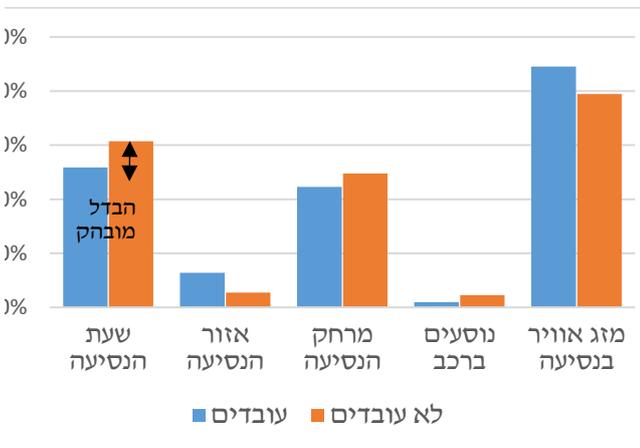

## 6.2 סיכום מודל העדפות

בבחינת העדפות הקשישים והדברים המשפיעים על נהיגתם נמצא כי שלושה אפיוני דרך חשובים עבורם: מזג האוויר נח בנסיעה, נסיעה בשעות היום ומרחק הנסיעה קצר יחסית. בהשוואה בין קבוצות דמוגרפיות שונות נמצא כי גורמים אלו חשובים לכל הקבוצות שנבדקו. יחד עם זאת, חשיבותן היחסית השתנתה מקבוצה לקבוצה. למרות שמשתנה מזג האוויר נמצא כמשפיע באופן החשוב ביותר, בקבוצות בהן ניתן להעריך כי קיים חוסר ביטחון יחסי בנהיגה – כקבוצת הנשים, קבוצת המבוגרים יותר, אלה הנוהגים בתדירות נמוכה או אלה שאינם עובדים - התעצמה באופן מובהק דווקא החשיבות של נסיעה בשעות היום ולא בשעות הלילה. חשיבותם של שאר הגורמים השתנתה אף היא אבל לא נמצאה מובהקת ברוב הקבוצות.

**לוח 6.4: סיכום מודל העדפות – השוואה בין קבוצות**

סימן חיובי: עלייה בחשיבות המשתנה בקבוצה המוגדרת לעומת הקבוצה השנייה.
סימן מינוס: ירידה בחשיבות המשתנה בקבוצה המוגדרת לעומת הקבוצה השנייה.
מסומן בעיגול: הבדל מובהק.

|  | מזג אוויר טוב | מלווה נוסף | מרחק נסיעה קצר | נסיעה עירונית | נסיעה ביום |
|---|---|---|---|---|---|
| מגדר - נשים | - | + | - | + | ⊕ |
| גיל – קשישים בני 75-80 | + | + | - | + | + |
| נוהגים בתדירות נמוכה | = | + | ⊕ | + | + |
| לא נשואים | ⊕ | + | - | + | ⊕ |
| לא גרים ליד הילדים | + | - | - | + | = |
| השכלה עד תיכונית | ⊖ | - | = | - | ⊕ |
| לא עובדים | - | + | + | - | ⊕ |

## 7. סיכום ומסקנות

שיעור הנהגים הקשישים מסך הנהגים עולה בהתמדה בעשורים האחרונים, עם עליית חלקם של הקשישים באוכלוסייה ועליית שיעור הנהגים בקרבם. מחקר זה בחן באמצעות סקר ייחודי את הרגלי הנהיגה של קשישים בגילאי 70 עד 80 בישראל בשנת 2016. נתוני הסקר וכן נתוני הלמ"ס מאפשרים לאפיין את הנהג הקשיש: כמחצית מהקשישים מעל גיל 65 מחזיקים ברישיון נהיגה. רוב הנהגים הקשישים הם גברים, ושיעור הנהגים בקרב הקשישים ובפרט שיעור הנהגות יורד עם עליית הגיל בקרב הקשישים. כ-90% מהנהגים הקשישים הם בעלי רשיון נהיגה מעל 40 שנה, ושיעור אלו המחזיקים גם רישיונות מיוחדים לרכב כבד, אוטובוס או מונית גבוה יחסית מהמאפיין את כלל הנהגים. מרבית הנהגים הקשישים נוהגים בתדירות גבוהה (מדי יום) כאשר תדירות הנסיעה נמוכה יותר בקרב נשים, עם עליית הגיל ובקרב לא נשואים.

שיעור תאונות הדרכים בהם מעורבים קשישים עולה לאורך השנים עם עליית משקלם של הנהגים הקשישים בכלל הנהגים. עם זאת, בקרב הקשישים עצמם שיעור תאונות הדרכים נמוך יחסית לנהגים צעירים יותר. במהלך ניתוח נתוני המחקר עלו חוסר ההתאמות המשולבות בתשובתם של הקשישים. רק אחוז נמוך יחסי של 13% אחוזים הודה שהיה מעורב בתאונת דרכים בעברו. אלה שזכרו מעורבות שכזו, זכרו אותה בעיקר מהחמש שנים האחרונות.

מחד, מצביעה הירידה בשיעור הנהגים הקשישים בקבוצות הגיל המבוגרות יותר, הירידה בתדירות הנסיעה ושיעור התאונות הנמוך יחסית המאפיין את אוכלוסיית הקשישים על רמת מודעות גבוהה של הקשיש המובילה אותו לפרוש מנהיגה בזמן או

להימנע מסוג של נהיגה היכולה לסכן אותו. מנגד, מצביע הסקר המיוחד שנערך במאמר על פער בין הרמה הממוצעת של איכות הנהיגה של הקשישים ובין רמת ההערכה העצמית הגבוהה בה העריכו הקשישים את רמת נהיגתם האישית.

הנתונים מראים כי שיעור הנפגעים בתאונות דרכים שהיה צריך להתקבל גבוה מזה עליו הצהירו הנשאלים. כמו כן כשליש מהנשאלים מודים כי קיבלו הערות מילדיהם לגבי הנהיגה שלהם, הם מודעים לכך שמגבלות פיזיות, טכניות או מנטליות, שהן מגבלות המאפיינות קשישים, פוגעות באיכות הנהיגה. הם מסכימים שיש פעולות במסגרת הנהיגה שהקשיש מתקשה איתם (כגון נסיעה ברוורס). עם זאת, הם מסרבים להכיר במשמעות התשובות שענו ביחס לכלל אוכלוסיית הקשישים וכן ביחס לעצמם כחלק מאוכלוסייה זו. אוסף הנקודות העולה מהשאלות הספציפיות לא מתחבר אצלם לתמונה השלמה ולמסקנות העולות מהן. הם אינם חושבים שצריך להגביל בצורה כלשהי נהיגת קשישים או לשנות את המבחנים בנושא זה, והם גם לא מתכוונים להפסיק לנהוג אלא אם כן מצבם הבריאותי לא יאפשר זאת. הם גם לא מכירים בכך שנהג קשיש שולט על הרכב פחות טוב, לא באופן כללי ולא ביחס לעצמם. עם זאת, הם מודים כי ייתכן שקשישים אינם מודעים למגבלות שלהם. התשובות מצביעות על רצון להרחקת כל חשש או ספק ביחס לנהיגתם והקשיים הכרוכים בה.

ממצאי סקר העדפות מצביעים על כך שמגבלות מזג אוויר ונסיעה בשעות היום הן שתי המגבלות המרכזיות איתן מתמודד הקשיש כאשר מגבלת הנהיגה בשעות החושך מחריפה ככל שהקבוצה הנבדקת נחשבת חלשה יותר מבחינת נהיגה, כלומר עם עליית שיעור הנשים, עליית הגיל או ירידת המיומנות של הנהג. שתי המגבלות הללו קשורות למצב הראיה אך במקרה של מזג אוויר גרוע הדבר קשור גם לשליטה בתנאים קשים ובלתי צפויים.

ניטור וסינון עצמי קיימים בנהיגת הקשישים, באופן עצמאי וביוזמת הקשיש וזאת גם ללא מגבלות חוקיות. נראה כי חלק מגברות הגיל היא היכולת לדעת מתי צריך "להיפרד מהמפתחות". עם זאת, ממחקר זה עולה כי הנהגים הקשישים הממשיכים לנהוג נוטים להפגין עודף אופטימיות לגבי מצבם ומרחיקים את עצמם מכל קישור למגבלה או לקושי העלול להיווצר או להיות כרוך בנהיגתם.

מדינת ישראל שינתה את הבדיקות הנדרשות ממחדשי רישיון הנהיגה בעשור האחרון. מחד, הקלה החקיקה על הנהגים הקשישים: מ-2010 חייב כל נהג בעל רשיון מגיל 70 בבדיקות רפואיות מידי 5 שנים ומגיל 80 מידי שנתיים, זאת לעומת מחויבות לבדיקה רפואית מגיל 65 כל שנתיים לפני שנת 2010. משנת 2016 חוקק כי חידוש הרישיון של בני 40+ יחייב את הנהג לעבור בדיקת ראייה.

אוכלוסיית הקשישים היא הטרוגנית מאוד וגם רמת הנהיגה של הפרטים בקרבה שונה מאחד לשני. השמירה על רמה גבוהה של עצמאות וחופש הכרוכים באפשרות הנהיגה חשובה מאוד לקשיש ולסביבתו. עם זאת, מחקר זה מעלה חששות בדבר רמת המודעות של הקשישים הנוהגים לגבי טיב הנהיגה שלהם. נראה כי לאור תוצאות המחקר נראה כי דווקא הדרישה להחמיר בבדיקות הראיה היא זו שהולכת לכיוון הנכון ולא ההקלה בבדיקות לבני ה-65 ומעלה. בשל הערכת היתר של הקשיש את מצבו והסיכון בנהיגתו יש לחזק את רמת הדיווחים בהם מחויבים הרופאים ולהגביר את מחויבותם בדווח ביחס למגבלות בנהיגת הקשישים המטופלים על ידם.

כלי נוסף אותו יש לשקול הוא אפשרות לפרישה הדרגתית של הקשישים מנהיגה על ידי מתן רישיונות מותנים לקשישים כשם שהם ניתנים לנהגים חדשים. המגבלה העיקרית העולה במחקר היא מגבלת הראיה משום כך יש לשקול מתן רישיונות שהם מותנים בתנאי מזג האוויר ובשעות האור. עבור הקשישים שהם מבוגרים מאוד (בני ה-85 ומעלה) יש לשקול הטלת מגבלות

חלקיות באופן גורף בשל עליית הסיכוי לקשיים פיזיים או מנטליים בקרב אוכלוסייה זו. עלייה ברמת הרגישות לנושא של הסביבה הקרובה, המקיפה את הקשיש חשובה גם היא מאוד במערך זה. בניית מערך הסברה הפונה לקשישים עצמם ולבני משפחותיהם גם הוא יכול לתת מענה למצב מורכב ורגיש זה.

| | שאלה | תשובה |
|---|---|---|
| א. האם שייך לאוכלוסיית הסקר | מהו גילך? | ___ |
| | האם יש לך רישיון נהיגה בתוקף ? | כן |
| | | לא |
| ב. מאפיינים דמוגרפים | מגדר | זכר |
| | | נקבה |
| | מהו המצב המשפחתי שלך? | נשוי/ה |
| | | גרוש/ה |
| | | אלמן/נה |
| | | רווק/ה |
| | האם יש לך ילדים? | כן |
| | | לא |
| | האם גר/ה לבד ? | כן |
| | | לא, עם בן/ת זוג |
| | | לא, עם הילדים |
| | | לא, בדיור מוגן |
| | האם מתגורר/ת בקרבת הילדים? | כן |
| | | לא |
| | השכלת המרואיין | לא למד כלל |
| | | בית ספר - כולל תיכון |
| | | ישיבה - מכל הסוגים |
| | | על תיכון |
| | | אוניברסיטה תואר אקדמי |
| | האת עובד? | כן |
| | | לא |
| | מהו היקף העבודה? | משרה מלאה |
| | | פעם פעמיים בשבוע |
| | | לעיתים רחוקות (פחות מפעם בשבוע) יותר |
| | מהו סטטוס המרואיין? | עצמאי |
| | | שכיר |
| | האם להגעה למקום העבודה יש צורך בנסיעה ? | כן |
| | | לא |
| | האם מתנדב/ת במסגרת כלשהי ? | כן, היכן? ____ |
| | | לא |
| | מהו היקף ההתנדבות? | כל יום |
| | | פעם פעמיים בשבוע |
| | | לעיתים רחוקות |
| | האם ההגעה למקום ההתנדבות מחייבת נסיעה? | כן |
| | | לא, קרוב למקום מגוריי |
| ג. מאפייני נהיגה | מהו סוג רישיון הנהיגה שלך ? | רכב פרטי |
| | | רכב כבד |
| | | מונית |
| | | אוטובוס |

| | שאלה | תשובה |
|---|---|---|
| | האם עומד לרשותך כלי רכב פרטי | כן, תמיד |
| | | כן, לעיתים רחוקות |
| | | לא, היה עד שנת |
| | | לא היה מעולם |
| | האם את/ה נוהג במהלך השבוע בכלי הרכב | כן, כל יום |
| | | כן, פעם פעמיים בשבוע |
| | | כן, לכל היותר פעם בשבוע |
| | | כן, לעיתים רחוקות |
| | | לא, רק מישהו אחר מהמשפחה נוהג/ת |
| התמודדות עם תקלות ברכב | כאשר יש תקר בגלגל | אני מתקן לבד |
| | | אני מזעיק קרובי משפחה |
| | | אני מזעיק שירות תיקונים |
| | | אחר |
| | כאשר יש מצבר "מת" | אני מתקן לבד |
| | | אני מזעיק קרובי משפחה |
| | | אני מזעיק שירות תיקונים |
| | | אחר |
| | כאשר יש תקלה אחרת ברכב | אני מתקן לבד |
| | | אני מזעיק קרובי משפחה |
| | | אני מזעיק שירות תיקונים |
| | | אחר |
| הצעות בני משפחה ביחד לנהיגה | האם המשפחה (הילדים) הציעו לך בזמן האחרון לצמצם את הנהיגה בלילה | כן |
| | | לא |
| | המשפחה (הילדים) הציעו לך בזמן האחרון להצטרף אליהם בנסיעה | כן |
| | | לא |
| | המשפחה (הילדים) הציעו לך בזמן האחרון להשתמש בתחבורה ציבורית | **כן** |
| | | לא |
| | המשפחה (הילדים) הציעו לך בזמן האחרון להצטרף לחברים כנוסע | כן |
| | | לא |
| | המשפחה (הילדים) הציעו לך בזמן האחרון לבדוק את כשירות נהיגתך | כן |
| | | כן |
| | המשפחה (הילדים) הציעו לך בזמן האחרון לחשוב על האפשרות לתלות את המפתחות | כן |
| | | לא |
| | המשפחה (הילדים) הציעו לך בזמן האחרון לא הציעו כל הצעה | כן |
| | | לא |
| מתי צריך להפסיק לנהוג? | צריך להגביל את נהיגת אנשים מבוגרים | לא |
| | | כן |
| | | כן, מגיל מסוים |
| | צריך להתנות את חידוש הרישיון במבחן כלשהו | לא |
| | | כן, איזה מבחן |
| | לנהוג כאשר יש בעיית שמיעה צריך להפסיק | כן |
| | | לא |
| | כאשר יש מוגבלות בהזזת הגוף | כן |
| | | לא |
| | כאשר יש מחלות לב | כן |
| | | לא |
| | כאשר יש ירידה בראייה | כן |
| | | לא |
| | מתי, לדעתך, תפסיק לנהוג ברכב שלך אחר | כן |
| | | לא |
| | האם יש לך חששות כלשהם מהנהיגה- | לא |
| | | כן, מהם |
| חשיבות מערכות הבטיחות במכונית | עד כמה חשובות לך המערכות הבאות? | |
| | מערכת למניעת החלקה ABS | חשוב מאוד |
| | | חשוב |
| | | ככה ככה |
| | | לא חשוב |
| | | מאוד לא חשוב |
| | מערכת לבדיקת לחץ אוויר בגלגלים | חשוב מאוד |
| | | חשוב |
| | | ככה ככה |
| | | לא חשוב |
| | | מאוד לא חשוב |
| | מערכת לשמירת מרחק ומניעת סטייה מהנתיב (AWACS) | חשוב מאוד |
| | | חשוב |
| | | ככה ככה |
| | | לא חשוב |
| | | מאוד לא חשוב |
| | אביזרי הבטיחות מצלמת רוורס | חשוב מאוד |
| | | חשוב |
| | | ככה ככה |
| | | לא חשוב |
| | | מאוד לא חשוב |
| | האם תהיה מוכן לשלם כסף על המערכות הבאות : | |
| | מערכת למניעת אובדן השליטה על הרכב | כן |
| | | לא |
| | מערכת למניעת החלקה ABS | כן |
| | | לא |
| | מערכת לבדיקת לחץ אוויר בגלגלים | **כן** |
| | | לא |
| | מערכת לשמירת מרחק ומניעת סטייה מהנתיב (AWACS) | כן |
| | | לא |
| | מצלמת רוורס | כן |
| | | כן |
| | לא מוכן להוסיף כסף על אף אחד מהם | כן |
| | | לא |
| תאונות דרכים | האם נפגעת אי פעם בתאונת דרכים | כן, תאונת דרכים ברכב-אני נהגתי. מתי? ___ |
| | | כן, תאונת דרכים ברכב-מישהו אחר נהג. מתי? ___ |
| | | כן, תאונת דרכים כהולך רגל. מתי |

| | | | |
|---|---|---|---|
| | | | לא |
| | תקופת התאונה | יום | |
| | | לילה | |
| | מקום התאונה | דרך עירונית | |
| | | דרך בינעירונית | |
| | היה מלווה בנסיעה | כן | |
| | | לא | |
| | סוג התאונה | קלה | |
| | | קשה | |
| | מי היה היה אשם | אני | |
| | | הנהג האחר | |
| | האם הייתה שלילה של רישיון הנהיגה | כן | |
| | | לא | |
| הסכמה עם היגדים | עד כמה אתה מסכים עם ההיגד הבא : | | |
| | לאנשים מבוגרים יש קושי בנהיגה לאחור - רוורס | לא מסכים כלל | |
| | | במידה מעטה מאד | |
| | | במידה מעטה | |
| | | במידה מסוימת | |
| | | במידה רבה | |
| | | במידה רבה מאד | |
| | אנשים מבוגרים משתמשים באופן מופרז בדוושת הבלמים | לא מסכים כלל | |
| | | במידה מעטה מאד | |
| | | במידה מעטה | |
| | | במידה מסוימת | |
| | | במידה רבה | |
| | | במידה רבה מאד | |
| | נהגים אחרים מרבים לצפור לאנשים מבוגרים | לא מסכים כלל | |
| | | במידה מעטה מאד | |
| | | במידה מעטה | |
| | | במידה מסוימת | |
| | | במידה רבה | |
| | | במידה רבה מאד | |
| | אנשים מבוגרים מתקשים יותר בחניה | לא מסכים כלל | |
| | | במידה מעטה מאד | |
| | | במידה מעטה | |
| | | במידה מסוימת | |
| | | במידה רבה | |
| | | במידה רבה מאד | |
| | אנשים מבוגרים כועסים ומתרגזים יותר בזמן הנהיגה | לא מסכים כלל | |
| | | במידה מעטה מאד | |
| | | במידה מעטה | |
| | | במידה מסוימת | |
| | | במידה רבה | |
| | | במידה רבה מאד | |
| | אנשים מבוגרים מתקשים לזהות תמרורים וסימנים בצדי הדרך | לא מסכים כלל | |
| | | במידה מעטה מאד | |

| | | |
|---|---|---|
| במידה מעטה | | |
| במידה מסוימת | | |
| במידה רבה | | |
| במידה רבה מאד | | |
| לא מסכים כלל | אנשים מבוגרים שולטים פחות טוב על הרכב בדרך כלל | |
| במידה מעטה מאד | | |
| במידה מעטה | | |
| במידה מסוימת | | |
| במידה רבה | | |
| במידה רבה מאד | | |
| לא מסכים כלל | אנשים מבוגרים מגיבים פחות טוב למצבים בלתי צפויים בדרך | |
| במידה מעטה מאד | | |
| במידה מעטה | | |
| במידה מסוימת | | |
| במידה רבה | | |
| במידה רבה מאד | | |
| לא מסכים כלל | אנשים מבוגרים לא מודעים תמיד למגבלות שלהם בקשר לנהיגה | |
| במידה מעטה מאד | | |
| במידה מעטה | | |
| במידה מסוימת | | |
| במידה רבה | | |
| במידה רבה מאד | | |

**נספח 2: תוצאות מודל העדפות בהשוואת תת קבוצות**

השוואה לפי מגדר:
לוח נ-2-1: מודל העדפות - גברים

| המאפיין | רמת המאפיין | מקדם | שגיאת תקן | T-TEST | חשיבות יחסית |
|---|---|---|---|---|---|
| שעת הנסיעה | נסיעה באור יום | 0.78043 | 0.0946 | 8.2496 | |
| | נסיעה בלילה | -0.78043 | 0.0946 | -8.2496 | 26.81% |
| אזור הנסיעה | נסיעה עירונית | 0.0794 | 0.08207 | 0.96745 | |
| | נסיעה בינעירונית | -0.0794 | 0.08207 | -0.96745 | 2.73% |
| מרחק הנסיעה | מרחק נסיעה ארוך | -0.83743 | 0.09496 | -8.81899 | |
| | מרחק נסיעה קצר | 0.83743 | 0.09496 | 8.81899 | 28.77% |
| נוסעים ברכב | נסיעה לבד | 0.01628 | 0.0821 | 0.19836 | |
| | נסיעה עם מישהו נוסף | -0.01628 | 0.0821 | -0.19836 | 0.56% |
| מזג אוויר בנסיעה | נסיעה במזג אוויר נח | 1.19752 | 0.10744 | 11.14645 | |
| | נסיעה במזג אוויר חורפי | -1.19752 | 0.10744 | -11.1465 | 41.14% |

לוח נ-2-2: מודל העדפות - נשים

| המאפיין | רמת המאפיין | מקדם | שגיאת תקן | T-TEST | חשיבות יחסית |
|---|---|---|---|---|---|
| שעת הנסיעה | נסיעה באור יום | 1.15753 | 0.10971 | 10.55058 | |
| | נסיעה בלילה | -1.15753 | 0.10971 | -10.5506 | 32.47% |
| אזור הנסיעה | נסיעה עירונית | 0.14431 | 0.08415 | 1.71492 | |
| | נסיעה בינעירונית | -0.14431 | 0.08415 | -1.71492 | 4.05% |
| מרחק הנסיעה | מרחק נסיעה ארוך | -0.76911 | 0.09865 | -7.79624 | |
| | מרחק נסיעה קצר | 0.76911 | 0.09865 | 7.79624 | 21.58% |
| נוסעים ברכב | נסיעה לבד | 0.1004 | 0.08402 | 1.1949 | |
| | נסיעה עם מישהו נוסף | -0.1004 | 0.08402 | -1.1949 | 2.82% |
| מזג אוויר בנסיעה | נסיעה במזג אוויר נח | 1.3934 | 0.11928 | 11.68131 | |
| | נסיעה במזג אוויר חורפי | -1.3934 | 0.11928 | -11.6813 | 39.09% |

לוח נ-2-3: השוואת המודלים לפי מגדר

| המאפיין | רמת המאפיין | גברים | | נשים | | T-TEST |
|---|---|---|---|---|---|---|
| | | מקדם | שגיאת תקן | מקדם | שגיאת תקן | |
| שעת הנסיעה | נסיעה באור יום | 0.78043 | 0.0946 | 1.15753 | 0.10971 | 2.60314 |
| | נסיעה בלילה | -0.78043 | 0.0946 | -1.15753 | 0.10971 | |
| אזור הנסיעה | נסיעה עירונית | 0.0794 | 0.08207 | 0.14431 | 0.08415 | 0.552216 |
| | נסיעה בינעירונית | -0.0794 | 0.08207 | -0.14431 | 0.08415 | |
| מרחק הנסיעה | מרחק נסיעה ארוך | -0.83743 | 0.09496 | -0.76911 | 0.09865 | 0.498949 |
| | מרחק נסיעה קצר | 0.83743 | 0.09496 | 0.76911 | 0.09865 | |
| נוסעים ברכב | נסיעה לבד | 0.01628 | 0.0821 | 0.1004 | 0.08402 | 0.716083 |
| | נסיעה עם מישהו נוסף | -0.01628 | 0.0821 | -0.1004 | 0.08402 | |
| מזג אוויר בנסיעה | נסיעה במזג אוויר נח | 1.19752 | 0.10744 | 1.3934 | 0.11928 | 1.22018 |
| | נסיעה במזג אוויר חורפי | -1.19752 | 0.10744 | -1.3934 | 0.11928 | |

השוואה לפי גיל:
לוח נ-2-4: מודל העדפות - גילאי 70-74

| המאפיין | רמת המאפיין | מקדם | שגיאת תקן | T-TEST | חשיבות יחסית |
|---|---|---|---|---|---|
| **שעת הנסיעה** | נסיעה באור יום | 0.96782 | 0.08016 | 12.07396 | |
| | נסיעה בלילה | -0.96782 | 0.08016 | -12.074 | 30.10% |
| **אזור הנסיעה** | נסיעה עירונית | 0.10809 | 0.06507 | 1.66121 | |
| | נסיעה בינעירונית | -0.10809 | 0.06507 | -1.66121 | 3.36% |
| **מרחק הנסיעה** | מרחק נסיעה ארוך | -0.83616 | 0.07699 | -10.8609 | |
| | מרחק נסיעה קצר | 0.83616 | 0.07699 | 10.86092 | 26.01% |
| **נוסעים ברכב** | נסיעה לבד | 0.02569 | 0.06488 | 0.39601 | |
| | נסיעה עם מישהו נוסף | -0.02569 | 0.06488 | -0.39601 | 0.80% |
| **מזג אוויר בנסיעה** | נסיעה במזג אויר נח | 1.2774 | 0.08861 | 14.41645 | |
| | נסיעה במזג אויר חורפי | -1.2774 | 0.08861 | -14.4165 | 39.73% |

לוח נ-2-5: מודל העדפות - גילאי 75-80

| המאפיין | רמת המאפיין | מקדם | שגיאת תקן | T-TEST | חשיבות יחסית |
|---|---|---|---|---|---|
| **שעת הנסיעה** | נסיעה באור יום | 0.94778 | 0.1527 | 6.20423 | |
| | נסיעה בלילה | -0.94778 | 0.1527 | -6.20423 | 31.09% |
| **אזור הנסיעה** | נסיעה עירונית | 0.11186 | 0.1288 | 0.86784 | |
| | נסיעה בינעירונית | -0.11186 | 0.1288 | -0.86784 | 3.67% |
| **מרחק הנסיעה** | מרחק נסיעה ארוך | -0.57417 | 0.1396 | -4.11133 | |
| | מרחק נסיעה קצר | 0.57417 | 0.1396 | 4.11133 | 18.83% |
| **נוסעים ברכב** | נסיעה לבד | 0.11987 | 0.1297 | 0.92421 | |
| | נסיעה עם מישהו נוסף | -0.11987 | 0.1297 | -0.92421 | 3.93% |
| **מזג אוויר בנסיעה** | נסיעה במזג אויר נח | 1.29502 | 0.1736 | 7.45782 | |
| | נסיעה במזג אויר חורפי | -1.29502 | 0.1736 | -7.45782 | 42.48% |

לוח נ-2-6: השוואה של המודלים לפי קבוצות גיל

| המאפיין | רמת המאפיין | גילאי 70-74 | | גילאי 75-80 | | T-TEST |
|---|---|---|---|---|---|---|
| | | מקדם | שגיאת תקן | מקדם | שגיאת תקן | |
| **שעת הנסיעה** | נסיעה באור יום | 0.96782 | 0.08016 | 0.94778 | 0.1527 | -0.1162 |
| | נסיעה בלילה | -0.96782 | 0.08016 | -0.94778 | 0.1527 | |
| **אזור הנסיעה** | נסיעה עירונית | 0.10809 | 0.06507 | 0.11186 | 0.1288 | 0.026125 |
| | נסיעה בינעירונית | -0.10809 | 0.06507 | -0.11186 | 0.1288 | |
| **מרחק הנסיעה** | מרחק נסיעה ארוך | -0.83616 | 0.07699 | -0.57417 | 0.1396 | 1.643366 |
| | מרחק נסיעה קצר | 0.83616 | 0.07699 | 0.57417 | 0.1396 | |
| **נוסעים ברכב** | נסיעה לבד | 0.02569 | 0.06488 | 0.11987 | 0.1297 | 0.649417 |
| | נסיעה עם מישהו נוסף | -0.02569 | 0.06488 | -0.11987 | 0.1297 | |
| **מזג אוויר בנסיעה** | נסיעה במזג אויר נח | 1.2774 | 0.08861 | 1.29502 | 0.1736 | 0.090402 |
| | נסיעה במזג אויר חורפי | -1.2774 | 0.08861 | -1.29502 | 0.1736 | |

השוואה לפי תדירות הנסיעה
לוח נ-2-7: מודל העדפות - נוהגים בתדירות גבוהה

| המאפיין | רמת המאפיין | מקדם | שגיאת תקן | T-TEST | חשיבות יחסית |
|---|---|---|---|---|---|
| שעת הנסיעה | נסיעה באור יום | 0.93465 | 0.08803 | 10.61725 | |
| | נסיעה בלילה | -0.93465 | 0.08803 | -10.6173 | 28.94% |
| אזור הנסיעה | נסיעה עירונית | 0.0944 | 0.07156 | 1.3192 | |
| | נסיעה בינעירונית | -0.0944 | 0.07156 | -1.3192 | 2.92% |
| מרחק הנסיעה | מרחק נסיעה ארוך | -0.90062 | 0.08656 | -10.4045 | |
| | מרחק נסיעה קצר | 0.90062 | 0.08656 | 10.40449 | 27.89% |
| נוסעים ברכב | נסיעה לבד | 0.00519 | 0.07143 | 0.07264 | |
| | נסיעה עם מישהו נוסף | -0.00519 | 0.07143 | -0.07264 | 0.16% |
| מזג אוויר בנסיעה | נסיעה במזג אוויר נוח | 1.29487 | 0.09867 | 13.12279 | |
| | נסיעה במזג אוויר חורפי | -1.29487 | 0.09867 | -13.1228 | 40.09% |

לוח נ-2-8: מודל העדפות לנוהגים בתדירות נמוכה יחסית

| המאפיין | רמת המאפיין | מקדם | שגיאת תקן | T-TEST | חשיבות יחסית |
|---|---|---|---|---|---|
| שעת הנסיעה | נסיעה באור יום | 1.06674 | 0.12511 | 8.52656 | 33.34% |
| | נסיעה בלילה | -1.06674 | 0.12511 | -8.52656 | |
| אזור הנסיעה | נסיעה עירונית | 0.13815 | 0.10345 | 1.33535 | 4.32% |
| | נסיעה בינעירונית | -0.13815 | 0.10345 | -1.33535 | |
| מרחק הנסיעה | מרחק נסיעה ארוך | -0.53787 | 0.11089 | -4.85036 | 16.81% |
| | מרחק נסיעה קצר | 0.53787 | 0.11089 | 4.85036 | |
| נוסעים ברכב | נסיעה לבד | 0.17561 | 0.10365 | 1.6942 | 5.49% |
| | נסיעה עם מישהו נוסף | -0.17561 | 0.10365 | -1.6942 | |
| מזג אוויר בנסיעה | נסיעה במזג אוויר נוח | 1.28078 | 0.13576 | 9.43445 | 40.04% |
| | נסיעה במזג אוויר חורפי | -1.28078 | 0.13576 | -9.43445 | |

לוח נ-2-9: השוואה בין המודלים לפי תדירות הנהיגה

| המאפיין | רמת המאפיין | נוהגים בתדירות גבוהה | | נוהגים בתדירות נמוכה יחסית | | T-TEST |
|---|---|---|---|---|---|---|
| | | מקדם | שגיאת תקן | מקדם | שגיאת תקן | |
| שעת הנסיעה | נסיעה באור יום | 0.93465 | 0.08803 | 1.06674 | 0.12511 | 0.863466 |
| | נסיעה בלילה | -0.93465 | 0.08803 | -1.06674 | 0.12511 | |
| אזור הנסיעה | נסיעה עירונית | 0.0944 | 0.07156 | 0.13815 | 0.10345 | 0.347806 |
| | נסיעה בינעירונית | -0.0944 | 0.07156 | -0.13815 | 0.10345 | |
| מרחק הנסיעה | מרחק נסיעה ארוך | -0.90062 | 0.08656 | -0.53787 | 0.11089 | 2.578654 |
| | מרחק נסיעה קצר | 0.90062 | 0.08656 | 0.53787 | 0.11089 | |
| נוסעים ברכב | נסיעה לבד | 0.00519 | 0.07143 | 0.17561 | 0.10365 | 1.353838 |
| | נסיעה עם מישהו נוסף | -0.00519 | 0.07143 | -0.17561 | 0.10365 | |
| מזג אוויר בנסיעה | נסיעה במזג אוויר נוח | 1.29487 | 0.09867 | 1.28078 | 0.13576 | -0.08395 |
| | נסיעה במזג אוויר חורפי | -1.29487 | 0.09867 | -1.28078 | 0.13576 | |

השוואה לפי מצב משפחתי:
לוח נ-2-10: מודל העדפות - נשואים

| המאפיין | רמת המאפיין | מקדם | שגיאת תקן | T-TEST | חשיבות יחסית |
|---|---|---|---|---|---|
| שעת הנסיעה | נסיעה באור יום | 0.87143 | 0.08007 | 10.88338 | |
| | נסיעה בלילה | -0.87143 | 0.08007 | -10.8834 | 29.73% |
| אזור הנסיעה | נסיעה עירונית | 0.07385 | 0.06691 | 1.10379 | |
| | נסיעה בינעירונית | -0.07385 | 0.06691 | -1.10379 | 2.52% |
| מרחק הנסיעה | מרחק נסיעה ארוך | -0.80146 | 0.07812 | -10.259 | |
| | מרחק נסיעה קצר | 0.80146 | 0.07812 | 10.25895 | 27.34% |
| נוסעים ברכב | נסיעה לבד | 0.02731 | 0.06677 | 0.409 | |
| | נסיעה עם מישהו נוסף | -0.02731 | 0.06677 | -0.409 | 0.93% |
| מזג אוויר בנסיעה | נסיעה במזג אוויר נוח | 1.15749 | 0.0879 | 13.16887 | |
| | נסיעה במזג אוויר חורפי | -1.15749 | 0.0879 | -13.1689 | 39.48% |

לוח נ-2-11: מודל העדפות - לא נשואים

| המאפיין | רמת המאפיין | מקדם | שגיאת תקן | T-TEST | חשיבות יחסית |
|---|---|---|---|---|---|
| שעת הנסיעה | נסיעה באור יום | 1.2263 | 0.1466 | 8.36475 | |
| | נסיעה בלילה | -1.2263 | 0.1466 | -8.36475 | 31.00% |
| אזור הנסיעה | נסיעה עירונית | 0.23989 | 0.11811 | 2.03102 | |
| | נסיעה בינעירונית | -0.23989 | 0.11811 | -2.03102 | 6.06% |
| מרחק הנסיעה | מרחק נסיעה ארוך | -0.72463 | 0.13373 | -5.41843 | |
| | מרחק נסיעה קצר | 0.72463 | 0.13373 | 5.41843 | 18.32% |
| נוסעים ברכב | נסיעה לבד | 0.10596 | 0.11701 | 0.90556 | |
| | נסיעה עם מישהו נוסף | -0.10596 | 0.11701 | -0.90556 | 2.68% |
| מזג אוויר בנסיעה | נסיעה במזג אוויר נוח | 1.65962 | 0.17207 | 9.64491 | |
| | נסיעה במזג אוויר חורפי | -1.65962 | 0.17207 | -9.64491 | 41.95% |

לוח נ-2-12: השוואה בין המודלים של נשואים ללא נשואים

| המאפיין | רמת המאפיין | נשואים | | לא נשואים | | T-TEST |
|---|---|---|---|---|---|---|
| | | מקדם | שגיאת תקן | מקדם | שגיאת תקן | |
| שעת הנסיעה | נסיעה באור יום | 0.87143 | 0.08007 | 1.2263 | 0.1466 | 2.124446 |
| | נסיעה בלילה | -0.87143 | 0.08007 | -1.2263 | 0.1466 | |
| אזור הנסיעה | נסיעה עירונית | 0.07385 | 0.06691 | 0.23989 | 0.11811 | 1.223169 |
| | נסיעה בינעירונית | -0.07385 | 0.06691 | -0.23989 | 0.11811 | |
| מרחק הנסיעה | מרחק נסיעה ארוך | -0.80146 | 0.07812 | -0.72463 | 0.13373 | 0.496076 |
| | מרחק נסיעה קצר | 0.80146 | 0.07812 | 0.72463 | 0.13373 | |
| נוסעים ברכב | נסיעה לבד | 0.02731 | 0.06677 | 0.10596 | 0.11701 | 0.583802 |
| | נסיעה עם מישהו נוסף | -0.02731 | 0.06677 | -0.10596 | 0.11701 | |
| מזג אוויר בנסיעה | נסיעה במזג אוויר נוח | 1.15749 | 0.0879 | 1.65962 | 0.17207 | 2.598729 |
| | נסיעה במזג אוויר חורפי | -1.15749 | 0.0879 | -1.65962 | 0.17207 | |

**השוואה לפי קירבה לבני משפחה:**

**לוח נ-2-13: מודל העדפות - מתגוררים בקרבת בני משפחה**

| המאפיין | רמת המאפיין | מקדם | שגיאת תקן | T-TEST | חשיבות יחסית |
|---|---|---|---|---|---|
| שעת הנסיעה | נסיעה באור יום | 0.97811 | 0.10106 | 9.67888 | |
| | נסיעה בלילה | -0.97811 | 0.10106 | -9.67888 | 29.83% |
| אזור הנסיעה | נסיעה עירונית | 0.10091 | 0.08163 | 1.2362 | |
| | נסיעה בינעירונית | -0.10091 | 0.08163 | -1.2362 | 3.08% |
| מרחק הנסיעה | מרחק נסיעה ארוך | -0.86768 | 0.09745 | -8.90377 | |
| | מרחק נסיעה קצר | 0.86768 | 0.09745 | 8.90377 | 26.46% |
| נוסעים ברכב | נסיעה לבד | 0.0639 | 0.08125 | 0.78654 | |
| | נסיעה עם מישהו נוסף | -0.0639 | 0.08125 | -0.78654 | 1.95% |
| מזג אוויר בנסיעה | נסיעה במזג אוויר נוח | 1.2683 | 0.11111 | 11.41435 | |
| | נסיעה במזג אוויר חורפי | -1.2683 | 0.11111 | -11.4144 | 38.68% |

**לוח נ-2-14: מודל העדפות - שאינם מתגוררים בקרבת בני משפחה**

| המאפיין | רמת המאפיין | מקדם | שגיאת תקן | T-TEST | חשיבות יחסית |
|---|---|---|---|---|---|
| שעת הנסיעה | נסיעה באור יום | 0.9452 | 0.10408 | 9.08132 | |
| | נסיעה בלילה | -0.9452 | 0.10408 | -9.08132 | 29.96% |
| אזור הנסיעה | נסיעה עירונית | 0.14822 | 0.08626 | 1.71824 | |
| | נסיעה בינעירונית | -0.14822 | 0.08626 | -1.71824 | 4.70% |
| מרחק הנסיעה | מרחק נסיעה ארוך | -0.73458 | 0.0983 | -7.4731 | |
| | מרחק נסיעה קצר | 0.73458 | 0.0983 | 7.4731 | 23.28% |
| נוסעים ברכב | נסיעה לבד | 0.0215 | 0.08609 | 0.24978 | |
| | נסיעה עם מישהו נוסף | -0.0215 | 0.08609 | -0.24978 | 0.68% |
| מזג אוויר בנסיעה | נסיעה במזג אוויר נוח | 1.30564 | 0.11767 | 11.0957 | |
| | נסיעה במזג אוויר חורפי | -1.30564 | 0.11767 | -11.0957 | 41.38% |

**לוח נ-2-15: השוואה בין המודלים לפי מגורים בקרבת בני משפחה**

| המאפיין | רמת המאפיין | השכלה עד תיכונית | | השוואה עובדים לא עובדים | | T-TEST |
|---|---|---|---|---|---|---|
| | | מקדם | שגיאת תקן | מקדם | שגיאת תקן | |
| שעת הנסיעה | נסיעה באור יום | 0.97811 | 0.10106 | 0.9452 | 0.10408 | -0.22685 |
| | נסיעה בלילה | -0.97811 | 0.10106 | -0.9452 | 0.10408 | |
| אזור הנסיעה | נסיעה עירונית | 0.10091 | 0.08163 | 0.14822 | 0.08626 | 0.398362 |
| | נסיעה בינעירונית | -0.10091 | 0.08163 | -0.14822 | 0.08626 | |
| מרחק הנסיעה | מרחק נסיעה ארוך | -0.86768 | 0.09745 | -0.73458 | 0.0983 | 0.961584 |
| | מרחק נסיעה קצר | 0.86768 | 0.09745 | 0.73458 | 0.0983 | |
| נוסעים ברכב | נסיעה לבד | 0.0639 | 0.08125 | 0.0215 | 0.08609 | -0.35818 |
| | נסיעה עם מישהו נוסף | -0.0639 | 0.08125 | -0.0215 | 0.08609 | |
| מזג אוויר בנסיעה | נסיעה במזג אוויר נוח | 1.2683 | 0.11111 | 1.30564 | 0.11767 | 0.230724 |
| | נסיעה במזג אוויר חורפי | -1.2683 | 0.11111 | -1.30564 | 0.11767 | |

השוואה לפי מצב תעסוקה:
לוח נ-2-16: עובדים

| המאפיין | רמת המאפיין | מקדם | שגיאת תקן | T-TEST | חשיבות יחסית |
|---|---|---|---|---|---|
| שעת הנסיעה | נסיעה באור יום | 0.66326 | 0.13487 | 4.91766 | |
| | נסיעה בלילה | -0.66326 | 0.13487 | -4.91766 | 25.85% |
| אזור הנסיעה | נסיעה עירונית | 0.16378 | 0.1236 | 1.32504 | |
| | נסיעה בינעירונית | -0.16378 | 0.1236 | -1.32504 | 6.38% |
| מרחק הנסיעה | מרחק נסיעה ארוך | -0.57144 | 0.13213 | -4.3248 | |
| | מרחק נסיעה קצר | 0.57144 | 0.13213 | 4.3248 | 22.28% |
| נוסעים ברכב | נסיעה לבד | -0.02481 | 0.12284 | -0.20196 | |
| | נסיעה עם מישהו נוסף | 0.02481 | 0.12284 | 0.20196 | 0.97% |
| מזג אוויר בנסיעה | נסיעה במזג אויר נוח | 1.14206 | 0.16091 | 7.09732 | |
| | נסיעה במזג אויר חורפי | -1.14206 | 0.16091 | -7.09732 | 44.52% |

לוח נ-2-17: לא עובדים

| המאפיין | רמת המאפיין | מקדם | שגיאת תקן | T-TEST | חשיבות יחסית |
|---|---|---|---|---|---|
| שעת הנסיעה | נסיעה באור יום | 1.06748 | 0.08545 | 12.49253 | |
| | נסיעה בלילה | -1.06748 | 0.08545 | -12.4925 | 30.74% |
| אזור הנסיעה | נסיעה עירונית | 0.09488 | 0.06672 | 1.4221 | |
| | נסיעה בינעירונית | -0.09488 | 0.06672 | -1.4221 | 2.73% |
| מרחק הנסיעה | מרחק נסיעה ארוך | -0.86079 | 0.08007 | -10.7501 | |
| | מרחק נסיעה קצר | 0.86079 | 0.08007 | 10.75006 | 24.79% |
| נוסעים ברכב | נסיעה לבד | 0.07911 | 0.06677 | 1.18491 | |
| | נסיעה עם מישהו נוסף | -0.07911 | 0.06677 | -1.18491 | 2.28% |
| מזג אוויר בנסיעה | נסיעה במזג אויר נוח | 1.37075 | 0.09481 | 14.45754 | |
| | נסיעה במזג אויר חורפי | -1.37075 | 0.09481 | -14.4575 | 39.47% |

לוח נ-2-18: השוואה בין המודלים של עובדים לשאינם עובדים

| המאפיין | רמת המאפיין | השכלה עד תיכונית | | עובדים לא עובדים | | השוואה |
|---|---|---|---|---|---|---|
| | | מקדם | שגיאת תקן | מקדם | שגיאת תקן | T-TEST |
| שעת הנסיעה | נסיעה באור יום | 0.66326 | 0.13487 | 1.06748 | 0.08545 | 2.53174 |
| | נסיעה בלילה | -0.66326 | 0.13487 | -1.06748 | 0.08545 | |
| אזור הנסיעה | נסיעה עירונית | 0.16378 | 0.1236 | 0.09488 | 0.06672 | -0.49054 |
| | נסיעה בינעירונית | -0.16378 | 0.1236 | -0.09488 | 0.06672 | |
| מרחק הנסיעה | מרחק נסיעה ארוך | -0.57144 | 0.13213 | -0.86079 | 0.08007 | -1.87284 |
| | מרחק נסיעה קצר | 0.57144 | 0.13213 | 0.86079 | 0.08007 | |
| נוסעים ברכב | נסיעה לבד | -0.02481 | 0.12284 | 0.07911 | 0.06677 | 0.743274 |
| | נסיעה עם מישהו נוסף | 0.02481 | 0.12284 | -0.07911 | 0.06677 | |
| מזג אוויר בנסיעה | נסיעה במזג אויר נוח | 1.14206 | 0.16091 | 1.37075 | 0.09481 | 1.224483 |
| | נסיעה במזג אויר חורפי | -1.14206 | 0.16091 | -1.37075 | 0.09481 | |

השוואה לפי השכלה

לוח נ-2-19: מודל העדפות - בעלי השכלה עד תיכונית

| המאפיין | רמת המאפיין | מקדם | שגיאת תקן | T-TEST | חשיבות יחסית |
|---|---|---|---|---|---|
| **שעת הנסיעה** | נסיעה באור יום | 0.5303 | 0.12279 | 4.31888 | |
| | נסיעה בלילה | -0.5303 | 0.12279 | -4.31888 | 21.77% |
| **אזור הנסיעה** | נסיעה עירונית | 0.17628 | 0.11618 | 1.51732 | |
| | נסיעה בינעירונית | -0.17628 | 0.11618 | -1.51732 | 7.24% |
| **מרחק הנסיעה** | מרחק נסיעה ארוך | -0.58012 | 0.12388 | -4.68279 | |
| | מרחק נסיעה קצר | 0.58012 | 0.12388 | 4.68279 | 23.82% |
| **נוסעים ברכב** | נסיעה לבד | 0.12714 | 0.1161 | 1.09502 | |
| | נסיעה עם מישהו נוסף | -0.12714 | 0.1161 | -1.09502 | 5.22% |
| **מזג אוויר בנסיעה** | נסיעה במזג אוויר נוח | 1.02193 | 0.14445 | 7.07473 | |
| | נסיעה במזג אוויר חורפי | -1.02193 | 0.14445 | -7.07473 | 41.96% |

לוח נ-2-20: מודל העדפות - בעלי השכלה על תיכונית לא אקדמאית

| המאפיין | רמת המאפיין | מקדם | שגיאת תקן | T-TEST | חשיבות יחסית |
|---|---|---|---|---|---|
| **שעת הנסיעה** | נסיעה באור יום | 1.20584 | 0.12549 | 9.60872 | |
| | נסיעה בלילה | -1.20584 | 0.12549 | -9.60872 | 31.67% |
| **אזור הנסיעה** | נסיעה עירונית | 0.06488 | 0.09216 | 0.70401 | |
| | נסיעה בינעירונית | -0.06488 | 0.09216 | -0.70401 | 1.70% |
| **מרחק הנסיעה** | מרחק נסיעה ארוך | -0.8278 | 0.1099 | -7.53245 | |
| | מרחק נסיעה קצר | 0.8278 | 0.1099 | 7.53245 | 21.74% |
| **נוסעים ברכב** | נסיעה לבד | 0.14502 | 0.09294 | 1.56032 | |
| | נסיעה עם מישהו נוסף | -0.14502 | 0.09294 | -1.56032 | 3.81% |
| **מזג אוויר בנסיעה** | נסיעה במזג אוויר נוח | 1.56454 | 0.14022 | 11.15807 | |
| | נסיעה במזג אוויר חורפי | -1.56454 | 0.14022 | -11.1581 | 41.08% |

לוח נ-2-21 מודל העדפות - בעלי השכלה אקדמאית

| המאפיין | רמת המאפיין | מקדם | שגיאת תקן | T-TEST | חשיבות יחסית |
|---|---|---|---|---|---|
| **שעת הנסיעה** | נסיעה באור יום | 1.17076 | 0.14108 | 8.29857 | |
| | נסיעה בלילה | -1.17076 | 0.14108 | -8.29857 | 30.46% |
| **אזור הנסיעה** | נסיעה עירונית | 0.15286 | 0.10734 | 1.42407 | |
| | נסיעה בינעירונית | -0.15286 | 0.10734 | -1.42407 | 3.98% |
| **מרחק הנסיעה** | מרחק נסיעה ארוך | -1.01885 | 0.13692 | -7.4412 | |
| | מרחק נסיעה קצר | 1.01885 | 0.13692 | 7.4412 | 26.51% |
| **נוסעים ברכב** | נסיעה לבד | -0.1091 | 0.10695 | -1.02014 | |
| | נסיעה עם מישהו נוסף | 0.1091 | 0.10695 | 1.02014 | 2.84% |
| **מזג אוויר בנסיעה** | נסיעה במזג אוויר נוח | 1.39147 | 0.15314 | 9.08618 | |
| | נסיעה במזג אוויר חורפי | -1.39147 | 0.15314 | -9.08618 | 36.21% |

לוח נ-2-22: השוואת המודלים לפי השכלה

| המאפיין | רמת המאפיין | השכלה עד תיכונית | | השכלה על תיכונית לא אקדמאית | | T-TEST השוואה לפי השכלה |
|---|---|---|---|---|---|---|
| | | מקדם | שגיאת תקן | מקדם | שגיאת תקן | |
| שעת הנסיעה | נסיעה באור יום | 0.5303 | 0.12279 | 1.20584 | 0.12549 | 3.847677 |
| | נסיעה בלילה | -0.5303 | 0.12279 | -1.20584 | 0.12549 | |
| אזור הנסיעה | נסיעה עירונית | 0.17628 | 0.11618 | 0.06488 | 0.09216 | -0.75121 |
| | נסיעה בינעירונית | -0.17628 | 0.11618 | -0.06488 | 0.09216 | |
| מרחק הנסיעה | מרחק נסיעה ארוך | -0.58012 | 0.12388 | -0.8278 | 0.1099 | -1.49563 |
| | מרחק נסיעה קצר | 0.58012 | 0.12388 | 0.8278 | 0.1099 | |
| נוסעים ברכב | נסיעה לבד | 0.12714 | 0.1161 | 0.14502 | 0.09294 | 0.120228 |
| | נסיעה עם מישהו נוסף | -0.12714 | 0.1161 | -0.14502 | 0.09294 | |
| מזג אוויר בנסיעה | נסיעה במזג אוויר נוח | 1.02193 | 0.14445 | 1.56454 | 0.14022 | 2.695337 |
| | נסיעה במזג אוויר חורפי | -1.02193 | 0.14445 | -1.56454 | 0.14022 | |